\documentclass[11pt,a4paper]{article}
\usepackage[T1]{fontenc}
\usepackage[utf8]{inputenc}

\usepackage{latexsym, amssymb, enumerate, amsmath, cite, tabu}
    \usepackage{graphicx}
\usepackage{color, cases,empheq}
\usepackage{authblk}

\newcommand{\ubar}{{\overline{u}}}
\newcommand{\vbar}{{\overline{v}}}

\newcommand{\pbar}{{\overline{p}}}
\newcommand{\velbar}{{\overline{\mathbf{v}}}}
\newcommand{\vel}{{\mathbf{v}}}

\newcommand{\Hbar}{{\bar{H}}}
\newcommand{\abar}{{\bar{a}}}
\newcommand{\Abar}{{\bar{A}}}
\newcommand{\sth}{{S^{\theta}}}
\newcommand{\sq}{{S^{q}}}

\newcommand{\tildeq}{{\tilde{Q}}}

\newcommand{\omjo}{{\omega_{\text{MJO}}}}

\newcommand{\obt}{{\omega_{\text{T}}}}

\newcommand{\odr}{{\omega_{\text{R}}}}
\newcommand{\okk}{{\omega_{\text{K}}}}

\newcommand{\oji}{{\omega_{1}}}
\newcommand{\ojt}{{\omega_{2}}}

\newcommand{\thbt}{{\theta_{\text{T}}}}

\newcommand{\tho}{{\theta_{0}}}
\newcommand{\thji}{{\theta_{1}}}
\newcommand{\thjt}{{\theta_{2}}}

\newcommand{\kbt}{{k_{\text{T}}}}

\newcommand{\kji}{{k_{1}}}
\newcommand{\kjt}{{k_{2}}}

\newcommand{\aji}{{\alpha_{1}}}
\newcommand{\ajt}{{\alpha_{2}}}

\newcommand{\lo}{{l^{(0)}}}
\newcommand{\lt}{{l^{(2)}}}
\newcommand{\ro}{{r^{(0)}}}
\newcommand{\rt}{{r^{(2)}}}
\newcommand{\vi}{{v^{(1)}}}
\newcommand{\qo}{{q^{(0)}}}
\newcommand{\qt}{{q^{(2)}}}
\newcommand{\ao}{{a^{(0)}}}

\newcommand{\AAA}{{\mathcal{A}}}
\newcommand{\CC}{{\mathcal{C}}}
\newcommand{\DD}{{\mathcal{D}}}
\newcommand{\NN}{{\mathcal{N}}}
\newcommand{\LL}{{\mathcal{L}}}
\newcommand{\LLU}{{\mathcal{L}_{\vec{U}}}}
\newcommand{\LLW}{{\mathcal{L}_{\vec{W}}}}
\newcommand{\FFU}{{\vec{F}_{\vec{U}}}}
\newcommand{\FFW}{{\vec{F}_{\vec{W}}}}

  \author[1]{Shengqian Chen \footnote{Corresponding author}}
  \author[2]{Andrew J. Majda}
  \author[3]{Samuel N. Stechmann} 
  \affil[1]{Department of Mathematics, University of Wisconsin -- Madison, Madison, Wisconsin, USA, sqchen@math.wisc.edu} 
  \affil[2]{Department of Mathematics, and Center for Atmosphere-Ocean Sciences, Courant Institute of Mathematical Science, New York University, New York, New York, USA, jonjon@cims.nyu.edu}
  \affil[3]{Department of Mathematics, and Department of Atmospheric and Oceanic Sciences, University of Wisconsin -- Madison, Madison, Wisconsin, USA, stechmann@wisc.edu}
  \title{Multiscale asymptotics for the Skeleton of the Madden-Julian Oscillation and Tropical--Extratropical Interactions}
  \date{}

\begin{document}
  \maketitle
  \abstract{
  A new model is derived and analyzed for tropical--extratropical
  interactions involving the Madden--Julian oscillation (MJO).
  The model combines (i) the tropical dynamics of the MJO and
  equatorial baroclinic waves and (ii) the dynamics of barotropic
  Rossby waves with significant extratropical structure,
  and the combined system has a conserved energy.
  The method of multiscale asymptotics is applied to systematically
  derive a system of ordinary differential equations (ODEs) for
  three-wave resonant interactions.  Two novel features are
  (i) a degenerate auxiliary problem with overdetermined equations
  due to a compatibility condition (meridional geostrophic balance)
  and (ii) cubic self-interaction terms that are not typically
  found in three-wave resonance ODEs.  Several examples illustrate
  applications to MJO initiation and termination, including cases of
  (i) the MJO, equatorial baroclinic Rossby waves, and barotropic
  Rossby waves interacting, and (ii) the MJO, baroclinic Kelvin waves, and
  barotropic Rossby waves interacting. 
%   Resonance with the Kelvin wave
%  is not possible in a dry model, but it occurs in the moist model
%  here through interactions with water vapor and convective activity.
%Resonance directly between the Kelvin wave and the barotropic Rossby wave is not possible,
%but it appears
% here through interactions with the MJO mode.
 Resonance with the Kelvin wave is not possible here
  if only dry variables are considered, 
  but it occurs in the moist model here through interactions with water vapor and convective activity.
  }
 % \keywords{tropical intraseasonal variability, tropical-extratropical interactions, multiscale asymptotic analysis}
  %\classification[PACS]{02.30.Mv, 92.60.Ox}
%  \communicated{communicated...}
%  \dedication{dedication...}
  %\received{First draft: June 11, 2015, second draft: October 12, 2015.}
 % \accepted{October 13, 2015.}
%  \journalyear{year...}
%  \journalvolume{volume..}
%  \journalissue{issue..}
  %\startpage{1}
%  \aop
%  \DOI{DOI...}
% \footnote{Corresponding author: Shengqian Chen, sqchen@math.wisc.edu. }

\section{Introduction}
The Madden-Julian Oscillation (MJO) is the dominant component of intraseasonal ($\approx$30-60~days)
variability in the tropics \cite{mj71, mj72, mj94}.
It is an equatorial wave envelope of complex multi-scale convective processes,
coupled with planetary-scale ($\approx$10,000-40,000~km) circulation anomalies.
Individual MJO events 
propagate eastward at a speed of roughly~5 m/s,
and their convective signal is most prominent over the
Indian and western Pacific Oceans \cite{z05}.
In addition to its significance to its own right,
the MJO also significantly affects many other components of the atmosphere-ocean-earth system,
such as monsoon development,
intraseasonal predictability in mid-latitude,
and the development of the El Ni$\tilde{\text{n}}$o southern oscillation (ENSO) \cite{lw11, z05}.

In addition to its strong tropical signal,
the MJO interacts with the global flow on the intraseaonsal timescales.
Teleconnection patterns between the global extratropics and the MJO 
were described in an early observational analysis 
by Weickmann (1983)~\cite{w83} and 
Weikmann \emph{et al.} (1985)~\cite{wetal85}.
Their results demonstrate coherent fluctuations between extratropical flow and
eastward-propagating outgoing longwave radiation (OLR) 
anomalies in the tropics. 
In the study by Matthews and Kiladis (1999)~\cite{mk99},
they illustrate the interplay between high-frequency transient 
extratropical waves and the MJO.
More recently, Weickmann and Berry (2009)~\cite{wb09} demonstrate that
convection in the MJO frequently evolves together with a portion 
of the activity in a global wind oscillation.

The interactions between extratropical waves and tropical convection 
have also been investigated in numerical models. 
To view the extratropical response to convective heating,
Jin and Hoskins (1995)~\cite{jh95} forced a primitive equation model 
with a fixed heat source in the tropics
in the presence of a climatological background flow
and obtain the Rossby wave train response in the result. 
To diagnose the more specific response to patterns of convection more like those of the observed
MJO, Matthews \emph{et al.} (2004)~\cite{metal04QJRMS} forced a primitive equation model in 
a climatological background flow with patterns of observed MJO. 
The resulting global response to that heating is similar in many respects to the observational analysis. 
The MJO initiation in response to extratropical waves was illustrated by Ray and Zhang (2010)~\cite{rz10}.
They show that a dry-channel model of the tropical atmosphere developed MJO-like signals in tropical wind fields
when forced by reanalysis fields at poleward boundaries.
Many other interesting studies on tropical--extratropical interactions have been carried out.
For example, see Lin \emph{et. al.} (2009)~\cite{lbd09} and Frederiksen and Frederiksen (1993)~\cite{ff93}  , and the review by Roundy (2011)~\cite{r11}.

How can one model the two-way interaction between MJO and extratropical waves in a simplified, integrative way? 
This is the primary question of the present paper.
Section~\ref{sec_model} introduces a planetary scale model 
for this purpose.
The model includes
(i) barotropic dynamics that span the tropics and extratropics, 
(ii) equatorial baroclinic dynamics,
and (iii) the interactive effects of moisture 
and convection. 
More specifically,
the model integrates the dry barotropic-first baroclinic interaction that has been studied 
by Majda and Biello (2003)~\cite{mb03} and Khouider and Majda (2005)\cite{km05tcfdi} 
with the MJO skeleton model first developed by Majda and Stechmann~(2009)~\cite{ms09pnas}
and further investigated by same authors~\cite{ms11}. 
The term ``skeleton'' used by the authors refers to the fundamental
features of the MJO on planetary/intraseasonal scales. 
In these previous studies,
Majda and Biello (2003) carry out a multiscale asymptotic analysis 
for the resonant interactions of
dry baroclinic Rossby waves and barotropic Rossby waves; and
the MJO skeleton model~\cite{ms09pnas} has captured three main features 
of the MJO on planetary/intraseasonal scales:
(i) slow eastward phase speed of roughly~5m/s;
(ii)  peculiar dispersion relation with $\mathrm{d}\omega/\mathrm{d}k \approx 0$;
(iii) horizontal quadrupole vortex structure.

A multiscale asymptotic analysis is presented here in Section~\ref{sec_asym}
and later sections,
adopting similar strategies as in Majda and Biello (2003),
based on long-wave scales, small amplitude assumption, and multiple long time scales.
The long-wave scaling leads to meridional geostrophic balance at the leading order,
which appears as a constraint in the partial differential equation (PDE) system.
This constraint complicates the auxiliary problem 
of the multiscale analysis.
In brief, the auxiliary problem is a necessary step for suppressing secular terms.
The complication here, which arises from the constraint,
is a degenerate operator,
which we rectify through a change of basis as described 
in Section~\ref{sec_aux_new}.
Section~\ref{sec_eig} presents the dispersion relation of the leading order
linear operator for both the baroclinic and barotropic modes.
In Section~\ref{sec_reduced}, 
by inspecting the dispersion curves of the linear operator,
three-wave resonant interactions are identified,
and an ODE system for wave interactions 
is derived using multiscale asymptotics.
A novel feature of this ODE system is the presence of
cubic self-interactions terms,
which are not typically found in three-wave interaction ODEs
(e.g., \cite{c85, m03, p87, r81, r82}).
Here the cubic self-interaction arises from the 
new nonlinearity in the MJO skeleton model
involving water vapor and convective activity.
Many previous studies have been presented
on resonant interactions of atmospheric waves,
and they usually focused on ``dry dynamics'' (e.g., \cite{fkn10, metal99, rd06, rd09, rtm08, rz06, rz07}).
See Khouider \emph{et. al.} (2013)~\cite{kms13} for a review.

Readers who are most interested in physical applications
can skip ahead to Section~\ref{sec_reduced},
where the reduced ODE model is presented.
The results with the model are then organized as follows.
First, in Section~\ref{sec_valid}, a validation study is presented
to explore the time scales of validity of the asymptotic model.
Second, in Section~\ref{sec_3wv_num}, numerical simulations are presented
for three-wave interactions.
Two cases of three-wave interactions are chosen for application
to MJO initiation and termination~\cite{rz10,swm15}
and tropical--extratropical interaction.
Each case therefore involves both the MJO and
barotropic Rossby waves, and the third wave is either
a dry baroclinic Rossby wave or a dry Kelvin wave.
Interactions involving only the Kelvin wave and the barotropic Rossby waves
are not possible, 
but by including the MJO mode from the model with water vapor and convection, 
the Kelvin waves are engaged in the resonance triad. 

% Here the resonant interaction of the
%baroclinic Kelvin wave and barotropic Rossby wave is
%made possible by using a model with
%water vapor and convective activity.

The details of the multiscale asymptotic analysis
are presented in the appendices.
In Appendix~\ref{sec_trun_sys} and~\ref{sec_trun_exp}, 
a meridional truncated system is formulated.
Appendix~\ref{sec_aux} provides the explicit 
formulation of the auxiliary problems.
Finally, the details of the derivation of the
reduced ODE system are given in Appendix~\ref{sec_multi}.

\section{The two-layer equatorial $\beta$-plane equations} \label{sec_model}

\begin{table}
  \centering 
  \begin{tabu}{llll}
%\tabucline[1pt]{-}
Par. & Derivation & Dim. val. & Description\\
\hline
$\beta$ & & $2.3\times10^{-11}$~m$^{-1}$s$^{-1}$ & Variation of Coriolis parameter with latitude\\
$\theta_0$ && $300$~K & Potential temperature at surface \\
$g$ && $9.8$~m~s$^{-2}$ & Gravitational acceleration\\
$H$ && $16$~km & Tropopause height\\
$N^2$ & $(g/\theta_0)\mathrm{d}\bar{\theta}/\mathrm{d}z$ & $10^{-4}$~s$^{-2}$ &
Buoyancy frequency squared\\
$c$ & $NH/\pi$ & $50$~m~s$^{-1}$ & Velocity scale\\
$X_e$ & $\sqrt{c/\beta}$ & $1500$~km & Equatorial length scale\\
$T$ & $X_e/c$ &$8$~hrs & Equatorial time scale\\
       & $HN^2\theta_0/(\pi g)$ & $15$~K & Potential temperature scale \\
       & $H/\pi$ & $5$~km & Vertical length scale\\
       & $H/(\pi T)$ & $0.2$~m~s$^{-1}$ & Vertical velocity scale \\
       & $c^2$ & $2500$~m$^2$~s$^{-2}$ & Pressure scale\\
 %   \tabucline[1pt]{-}
\end{tabu}
  \caption{Constants and reference scales for nondimensionalization.}
  \label{tb_nondim}
\end{table}

The nondimensional two-layer equatorial $\beta$-plane equations for the barotropic and 
baroclinic MJO skeleton model are given by 
\begin{subequations}
\begin{align}
&\frac{\partial \velbar}{\partial t} + \velbar \cdot \nabla \velbar +y \velbar^{\perp}+ \nabla \pbar =-\frac{1}{2} \nabla \cdot (\vel \otimes \vel), \label{eq_velb} \\
&\nabla \cdot \velbar =0, \label{div_free}\\
&\frac{\partial \vel}{\partial t} + \velbar \cdot \nabla \vel - \nabla \theta + y \vel^{\perp}=-\vel \cdot \nabla \velbar, \label{eq_vel}\\
&\frac{\partial \theta}{\partial t} + \velbar \cdot \nabla \theta - \nabla \cdot \vel = \delta^2( \Hbar a - \sth) , \\
&\frac{\partial q}{\partial t} + \velbar \cdot \nabla q + \tilde{Q}\nabla \cdot \vel =-\delta^2(\Hbar a -\sq), \\
& \frac{\partial a}{ \partial t}  = \Gamma q a. \label{eq_a}
\end{align}
\label{eq_full}
\end{subequations}
These equations combine the MJO skeleton model 
\cite{ms09pnas}
and nonlinear interactions between the baroclinic and barotropic modes
\cite{mb03}.
The equations have been nondimensionalized using
the scales listed in Table~\ref{tb_nondim}.
Here $\velbar = (\ubar,\vbar)$ and $\pbar$ are barotropic velocity and pressure;
$\vel=(u,v)$ and $\theta$ are baroclinic velocity and potential temperature; and $q$ is water vapor (sometimes referred to as ``moisture''). 
The tropical convective activity envelope is denoted by $\delta^2 a$,
where $\delta$ is a small parameter that modulates the scales of tropical convection envelope.
Likewise, $\delta^2 $ is also applied to quantities $\sth$ and $\sq$,
radiative cooling and the moisture source.
In this paper, $\delta^2\sth$ and $\delta^2\sq$ are set to be constants for energy conservation,
although they usually have both spatial and temporal variance in reality.
Together with $\Gamma$, $\Hbar$ and $\tildeq$, 
the coefficients are described in table~\ref{tb_coef}.

\begin{table}
  \centering 
  \begin{tabu}{llll}
%\tabucline[1pt]{-}
Par. & Non-dim. val. & Dim. val. & Description\\
\hline
   $\Gamma$&  1  & $\sim$0.5 /day/(g/kg)  & Convective growth/decay rate \\
   $\Hbar$& 0.23 & $\sim$10~K/day& Parameter to rescale $a$ \\
   $\delta^2\abar$ & $\delta^2$  &  & Convective activity envelope at RCE state\\
  $\tildeq$ &  0.9 &   & Non-dim. background vertical moisture gradient \\
  $\delta^2\sth$    &  $\delta^2\Hbar$ & & Radiative cooling rate \\
    $\delta^2 \sq$    &  $\delta^2\Hbar$ &   & Moisture source\\
 %   \tabucline[1pt]{-}
\end{tabu}
  \caption{Parameters of the MJO skeleton model, and relation to small parameter $\delta$.}
  \label{tb_coef}
\end{table}
In equations~(\ref{eq_full}), $\vel^{\perp} = (-v,u)$ is the $\beta$-plane approximation
of tropical Coriolis force,
and $x$ and $y$ denote the zonal and meridional coordinates.
Without the moisture $q$ and convection envelope $a$, 
system~(\ref{eq_full}) is the two-vertical-mode Galerkin truncation for the
Boussinesq equations with rigid-lid boundary on the top and bottom of the domain (see e.g., \cite{nz00,ms01b,mb03, km05tcfdi})
Without the barotropic wind $\velbar$, the system is the MJO skeleton model
on the first baroclinic mode 
before passing to the long wave limit (see e.g., \cite{ms09pnas, ms11}).

Note that for the primitive equations~(\ref{eq_full}), 
a total energy is conserved, and it is composed of four parts, 
dry barotropic energy $\mathcal{E}_T$,
dry baroclinic energy $\mathcal{E}_C$,
moisture energy $\mathcal{E}_M$,
and convective energy $\mathcal{E}_A$:
\begin{subequations}
\begin{align}
\mathcal{E}_T(t) & = \frac{1}{2}\int_{-Y}^{Y}\int_{0}^{X}|\velbar|^2\mathrm{d}x\mathrm{d}y\\
\mathcal{E}_C(t) & = \frac{1}{4}\int_{-Y}^{Y}\int_{0}^{X}|\vel|^2+\theta^2 \mathrm{d}x\mathrm{d}y\\
\mathcal{E}_M(t) & = \frac{1}{4}\int_{-Y}^{Y}\int_{0}^{X} \frac{1}{\tildeq(1-\tildeq)}(q+\tildeq \theta)^2\mathrm{d}x\mathrm{d}y\\
\mathcal{E}_A(t) & = \frac{\delta^2}{2}\int_{-Y}^{Y}\int_{0}^{X}\frac{1}{\tildeq \Gamma}\left[\Hbar a-\sth \log(a)\right]\mathrm{d}x\mathrm{d}y
\end{align}
\end{subequations}
This energy conservation forms the design principle as appeared in \cite{ms09pnas, mb03}.
Note that in (\ref{eq_a}), the conserved quantity $\mathcal{E}$ still holds
if $a$ is also advected by the barotropic wind, i.e., with an additional term $\ubar \cdot \nabla a$ in (\ref{eq_a}).
But we do not include this term because later, when meridional truncation is applied to the system,
the energy conservation will not hold with this additional term.

Using the streamfunction $\psi$ for barotropic mode, which satisfies $(\ubar,\vbar) =(-\psi_y, \psi_x)$,
the barotropic equation can also be written as 
\begin{equation}
 \frac{D}{Dt} \Delta \psi + \psi_x + \frac{1}{2} \nabla \cdot \left[ - (\vel u)_y + (\vel v)_x\right] = 0,
 \label{eq_bt_psi}
 \end{equation}
 where 
 $$ \frac{D}{Dt}  =  \frac{\partial}{\partial t} + \ubar  \frac{\partial}{\partial x} +  \vbar  \frac{\partial}{\partial y}$$
 represents advection by the barotropic wind.

\subsection{The zonal-long wave scaled model}

To consider the planetary scale of the coupled equations in (\ref{eq_bt_psi}) and (\ref{eq_vel})-(\ref{eq_a}),
zonal variations are assumed to depend on a longer scale,
as are temporal variations.
The long zonal and long temporal coordinates are defined as
\begin{equation}
x' = \delta x, \qquad t' = \delta t.
\end{equation}
Correspondingly, the meridional velocity is also scaled so that
\begin{equation}
v = \delta v'.
\end{equation}
The equations in (\ref{eq_bt_psi}) and (\ref{eq_vel})-(\ref{eq_a}) become
the long-wave-scaled system:
\begin{subequations}
\begin{align}
& \frac{D}{Dt'} \psi_{yy} + \psi_{x'} - \frac{1}{2} \nabla' \cdot \left[ (\vel' u)_y\right] + 
\delta^2 \left\{ \frac{D}{Dt'} \psi_{x'x'}+\frac{1}{2} \nabla' \cdot \left[ (\vel' v')_{x'} \right] \right\} = 0,\\
&\frac{D}{Dt'}u- \theta_x - y u  -\vel' \cdot \nabla' \psi_y = 0, \label{eq_long_}\\
& -\theta_y + y u + \delta^2\left(\frac{D}{Dt'} v' + v'\cdot \nabla ' \psi_{x'}\right)=0 \\
&\frac{D}{Dt'}\theta  - \nabla' \cdot \vel' - \delta ( \Hbar a - \sth)=0 , \label{eq_long_th} \\
&\frac{D}{Dt'}q + \tilde{Q}\nabla' \cdot \vel' + \delta (\Hbar a -\sq)=0 , \label{eq_long_q} \\
& \delta\frac{\partial }{ \partial t'}a  - \Gamma q a=0.  \label{eq_long_a}
\end{align}
\label{eq_long}
\end{subequations}
Here the primes represent the long-wave scaled coordinates and variables:
$$\frac{D}{Dt'} = \frac{\partial }{ \partial t'} + u \frac{\partial }{ \partial x'}+v' \frac{\partial }{ \partial y'},
\quad
\nabla' = (\frac{\partial }{ \partial x'}, \frac{\partial }{ \partial y}), \quad
\text{and}\quad  \velbar' = (u, v').
 $$
System (\ref{eq_long}) is in the same form as in \cite{mb03} when
the moisture and convection are neglected. 

Furthermore, the convective activity $a$ will be written as
an anomaly from the state of
radiative-convective equilibrium (RCE):
$$\sth = \sq = \Hbar \abar, $$ 
where these constants take the values given in table~\ref{tb_coef}.
When $a=\abar$, the external radiative cooling/moisture source
is in balance with the source/sink from the convection, and 
the system achieves RCE.
By writing $a$ as an anomaly with respect to the RCE value $\bar{a}$,
the forcing terms in 
(\ref{eq_long_th}) and (\ref{eq_long_q}) can be written as
\begin{equation}
\Hbar a - \sth = \Hbar a - \sq =  \Hbar (a - \abar) = \Hbar a',
\end{equation}
and the equation for the convection envelope (\ref{eq_long_a}) can be written as
\begin{equation}
\frac{\partial }{ \partial t'}a'  - \Gamma \abar q   =\Gamma a' q.
\end{equation}

\subsection{Long time scales for tropical--extratropical interactions}
 
The small parameter $\delta$ is also used for introducing 
 two longer time scales:
 $$T_1 = \delta t', \quad T_2 = \delta^2 t'.$$
Their units match the units of intraseasonal timescales as appeared in \cite{mb03, ms09cmtJAS}.

In the next section,  the system is expanded by matching the orders of $\delta$.

\section{Asymptotic expansions for the interaction of barotropic and equatorial baroclinic waves}\label{sec_asym}

In this section, asymptotic expansions are carried out twice:
first for the 2-D system as functions of $x$ and $y$,
and second for a system using a truncated basis 
for variations in the $y$ direction.
The first expansion (2-D) provides the clearest presentation,
but the 2-D linear operator does not have eigenvalues and
eigenfunctions that are easily accessible.
For this reason, the second, truncated system is introduced
and provides a linear operator with known eigenvalues
and eigenfunctions.

\subsection{The 2-D equations} \label{sec_2d}

Assume that 
the solutions have an asymptotic structure with \textbf{ansatz}:
\begin{subequations}
\begin{align}
(\psi, u, v', \theta, q)&= \delta^2(\psi_1, u_1, v_1, \theta_1,q_1)+ \delta^3 (\psi_2, u_2, v_2, \theta_2,q_2)
\nonumber\\
&+\delta^4 (\psi_3, u_3, v_3, \theta_3,q_3) + O(\delta^5) \\
%& (\psi, l, r, v', q) = \nonumber\\
%&\qquad\qquad \delta^2(\psi_1, l_1, r_1, v_1, q_1)+ \delta^3(\psi_2, l_2, r_2, v_2, q_2)
%+ \delta^4(\psi_3, l_3, r_3, v_3, q_3)+O(\delta^5), \\
 \text{and }\qquad \qquad a' &= \delta a_1 + \delta^2 a_2 + \delta^3 a_3 + O(\delta^4),
\end{align}
\label{assume}
\end{subequations}
This small amplitude assumption is consistent with
a small Froude number assumption as
in~\cite{mb03}.
All variables are assumed to depend on three time scales:
$t'$ and $T_1=\delta t'$ and $T_2=\delta^2 t'$.
The system is expanded over three orders of magnitude. 

The \textbf{first order system} is
\begin{subequations}
\begin{align}
& \psi_{1y'y't'} + \psi_{1x'} =0,\\
& u_{1t'}-\theta_{1x'}-y'v_1 =0,\\
& -\theta_{1y'}+y'u_1=0,\\
& \theta_{1t'} - u_{1x'}-v_{1y'}-\Hbar a_1=0,\\
& q_{1t'}+\tilde{Q}(u_{1x'}+v_{1y'}) +\Hbar a_1=0,\\
& a_{1t'} - \Gamma \abar q_1=0 .
\end{align}
\label{asym_2d_1}
\end{subequations}
This system defines the leading-order linear operator,
including a combination of barotropic Rossby waves
\cite{mb03} from the dynamics of $\psi$
and baroclinic equatorial long-waves including the MJO
\cite{ms09pnas} from the other variables.

The \textbf{second order system} is
\begin{subequations}
\begin{align}
& \psi_{2y'y't'} + \psi_{2x'} =-\psi_{1y'y'T_1},\\
& u_{2t'}-\theta_{2x'}-y'v_2 =-u_{1T_1},\\
& -\theta_{2y'}+y'u_2=0,\\
& \theta_{2t'} - u_{2x'}-v_{2y'}-\Hbar a_2 =-\theta_{1T_1} ,\\
& q_{2t'}+\tilde{Q}(u_{2x'}+v_{2y'})+\Hbar a_2  = -q_{1T_1},\\
& a_{2t'} -\Gamma \abar q_2= -a_{1T_1} + \Gamma a_1q_1.
\end{align}
\label{asym_2d_2}
\end{subequations}
Note that
the only nonlinear term is from the  $q$-$a$ interaction: $\Gamma a_1q_1$.
 
 The \textbf{third order system} is
\begin{subequations}
\begin{align}
& \psi_{3y'y't'} + \psi_{3x'} =-\psi_{1y'y'T_2}-\psi_{2y'y'T_1}\nonumber\\
&\qquad -\left[ \psi_{1x'x't'}+\psi_{1x'}\psi_{1y'y'y'}-\psi_{1y'}\psi_{1x'y'y'}-\frac{1}{2}(u_1^2)_{x'y'}-\frac{1}{2}(u_1v_1)_{y'y'} \right],\\
& u_{3t'}-\theta_{3x'}-y'v_3 =-u_{1T_2}-u_{2T_1}-\left[\psi_{1x'}u_{1y'}-\psi_{1y'}u_{1x'} - u_1\psi_{1x'y'}-v_1\psi_{1y'y'} \right],\\
& -\theta_{3y'}+y'u_3=-v_{1t'},\\
& \theta_{3t'} - u_{3x'}-v_{3y'}-\Hbar a_3=-\theta_{1T_2}-\theta_{2T_1} -\psi_{1x'}\theta_{1y'}+\psi_{1y'}\theta_{1x'} ,\\
& q_{3t'}+\tilde{Q}(u_{3x'}+v_{3y'}) +\Hbar a_3= -q_{1T_2}-q_{2T_1}-\psi_{1x'}q_{1y'}+\psi_{1y'}q_{1x'},\\
& a_{3t'}- \Gamma \abar q_3  =-a_{1T_2}-a_{2T_1}+ \Gamma (a_1q_2+a_2q_1)-\psi_{1x'}a_{1y'}+\psi_{1y'}a_{1x'}.
\end{align}
\label{asym_2d_3}
\end{subequations}
Note that more nonlinear terms arise,
including
the baroclinic--barotropic interactions as in~\cite{mb03}.

The total energy is approximately conserved:
\begin{equation}
\frac{\mathrm{d}}{\mathrm{d}t'} \mathcal{E}^{(3)} = o(\delta^4).
\end{equation}
 from the asymptotic expansions
 (\ref{asym_2d_1})-(\ref{asym_2d_3}),
 where 
 $\mathcal{E}^{(3)}$ is the energy of the variables
 from the leading 3 orders of magnitude from the ansatz in (\ref{assume}). 
This approximated energy conservation is important to preserve
in future derivations. 

The asymptotic expansions (\ref{asym_2d_1})-(\ref{asym_2d_3})
are analogous to the expansions in~\cite{mb03},
where the system did not have moisture $q$ nor convection envelope $a$.
Nonetheless, there are essential differences described
in table~\ref{tb_model_comp}.
For example, in~\cite{mb03}, the small parameter was
$\epsilon=\delta^2$, in which case the ansatz in (\ref{assume})
includes only the $\delta^2$ and $\delta^4$ terms, and
only the systems in
(\ref{asym_2d_1}) and (\ref{asym_2d_3}) arise.
On the other hand, here the small parameter is $\delta$,
which leads to the additional $\delta^3$ terms in the ansatz (\ref{assume})
and the additional system in (\ref{asym_2d_2}).

Unfortunately, the 2-D linear operator defined in (\ref{asym_2d_1})
does not have eigenvalues and eigenfunctions that are easily accessible.
This is due to the effects of moisture $q$ and convective activity $a$,
since on the other hand the ``dry'' dynamics without $q$ and $a$ do have
well-known eigenvalues and eigenfunctions
\cite{mb03}.
Since the method of multiscale asymptotics relies heavily on the
leading order linear operator, further progress cannot be made
with the systems (\ref{asym_2d_1})--(\ref{asym_2d_3}).

To circumvent this issue, a truncated basis will be introduced next
in section \ref{sec_merid_trunc} for variations in the $y$ direction, and
the truncated system has eigenvalues and eigenvectors
that have previously been presented
\cite{ms09pnas}.
The truncated system utilizes Riemann invariants
$l=-(u+\theta)/\sqrt{2}$ and $r=(u-\theta)/\sqrt{2}$
\cite{m03}
in place of the variables $u$ and $\theta$.

\begin{table}
  \centering 
   \begin{tabu}{lcl}
   %\tabucline[1pt]{-}
%  \begin{tabu}{L{.48\hsize}C{.02\hsize}L{.48\hsize}}
% after \\ : \hline or \cline{col1-col2} \cline{col3-col4} ...
   \centering{MJO skeleton model } && \centering{Dry wave model in~\cite{mb03}}  \\
   \tabucline[1.5pt]{-}
   Moisture equations involving $q$ and $a$ are included&&
   Dry dynamics only\\
\hline
Small parameter $\delta$ &&
Small parameter $\epsilon = \delta^2$ \\
\hline
 $\delta = \sqrt{0.1}$, $\epsilon = 0.1$
 &&
 \\
 \hline
Two long time scales: $T_1 = \delta t'$ and $T_2 = \delta^2 t'$ &&
One long time scale: $\tau = \epsilon t'$ \\
\hline
$T_1 = O(3 \text{ day})$, $T_2 = O(10\text{ day})$ 
&&
\\
\hline
Linear system is dispersive &&
Linear system is non-dispersive\\
\end{tabu}
  \caption{Comparisons between the MJO model here and the dry wave model of~\cite{mb03}.}
  \label{tb_model_comp}
\end{table}

\subsection{Meridional truncated system}
\label{sec_merid_trunc}
To simplify the system,
a meridional truncation is adopted as in the truncations 
in MJO skeleton model~\cite{ms09pnas} by using parabolic cylinder functions
for baroclinic variables.
The barotropic variables are assumed to be sinusoidal as in~\cite{mb03}.
Although mostly similar to~\cite{ms09pnas} and~\cite{mb03}, some modifications are needed for energy conservation, as described further below. 
Explicitly, the \textbf{meridional truncation} is
\begin{subequations}
\begin{align}
 \psi &= B(x,t)\sin(Ly)\label{eq_psi_trun}\\
 (l, r, q) &=\left(\lo (x,t),\ro(x,t),\qo (x,t)\right)\Phi_0(y), \\
&+ \left(\lt (x,t),\rt (x,t),\qt(x,t)\right)\Phi_2(y), \\
 v'& = \vi \Phi_1(y),\\
 a' &= \ao(x,t)\Phi_0(y),
\end{align}
\label{merid_trunc}
\end{subequations}
where $\Phi_m(y)$ are parabolic cylinder functions:
\begin{equation}
\Phi_m(y) = \left(m! \sqrt{\pi}\right)^{-\frac{1}{2}} 2^{-\frac{m}{2}} e^{-\frac{y^2}{2}}H_m(y),
\end{equation}
with Hermite polynomials $H_m(y)$ defined by
\begin{equation}
H_m(y) = (-1)^m e^{y^2}\frac{d^m e^{-y^2}}{dy^m}.
\end{equation}
The parabolic cylinder functions form an orthonormal basis on the 1D function space.
The first few functions are
\begin{equation}
\Phi_0(y) = {\pi^{-\frac{1}{4}}} e^{-y^2/2},\quad
\Phi_1(y) ={\pi^{-\frac{1}{4}}}  {\sqrt{2}} y e^{-y^2/2},\quad
\Phi_2(y) = {\pi^{-\frac{1}{4}}}  \frac{1}{\sqrt{2}}  (2y^2-1)e^{-y^2/2}.
\end{equation}
The parabolic cylinder functions satisfy the following identities:
\begin{equation}
\LL_+ \Phi_m(y) = (2m )^{1/2} \Phi_{m-1}(y), \quad \LL_- \Phi_m(y) = -\left[2(m+1)\right]^{1/2} \Phi_{m+1}(y),
\end{equation}
which help to simplify the expression, where
the operators $\LL_{\pm}$ are defined as
$$\LL_{\pm} = \frac{\partial}{\partial y} \pm y .$$
In the truncation~(\ref{eq_psi_trun}), the parameter
 $$L = \frac{2\pi}{2Y}$$
  is the meridional wavenumber, and $2Y$ is the meridional wavelength.

Applying this meridional truncation to the 2-D model in (\ref{eq_long})
leads to the \textbf{truncated system}:
\begin{subequations}
\begin{align}
& L^2 YB_{t'} - Y B_{x'}= \CC_{\text{T}}+\delta^2 \left(Y B_{x'x't'}+\DD_{\text{T}}\right), \label{eq_trunc_barotropic} \\
& \NN \vec{U}_t + \LLU \vec{U} = \vec{\CC}_{\text{C}} + \delta^2 \vec{\DD}_{\text{C}},
\end{align}
\label{eq_trunc}
\end{subequations}
where $\vec{U} = ( \lo,\lt,\ro,\rt,\vi', \qo,\qt, \ao)$,
$\LLU$ is the spatial linear operator, and
$\NN = \text{diag}(1,1,1,1,0,1,1,1)$ is an $8\times8$ diagonal matrix
with the $0$ placed in the $\vi'$ element of the diagonal.
At the right hand side, $\CC_{\text{T}}$, $\DD_{\text{T}}$, 
$ \vec{\CC}_{\text{C}}$ and $ \vec{\DD}_{\text{C}}$ are bilinear terms at different orders 
in the long-wave scaled system. 
These bilinear terms are from advection terms in the 2-D system 
(i.e., $\velbar \cdot \nabla \velbar$, $\nabla \cdot (\vel \otimes \vel)$, $\velbar \cdot \nabla \vel$,
$\vel \cdot \nabla \velbar$) and nonlinear $q$-$a$ interaction: $\Gamma q a$.
The detailed expressions are given in (\ref{eq_merid}) in Appendix~\ref{sec_trun_sys}.

This particular meridional truncation (\ref{merid_trunc}) is used 
because it maintains the $\lo$-$\vi$-$\rt$ triplets to obtain the correct dispersion
relation for baroclinic Rossby waves~\cite{m03}.
Also, this meridional truncation imitates the energy conservation as in the full system
with minimal alteration in the $\ao$ equation.
Other truncations may also be considered,
such as neglecting $\lt$ and $\qt$ as in MJO skeleton model~\cite{ms09pnas},
 and we have explored several other options.
However, other truncations can potentially contribute to energy imbalance from the barotropic advection acting
on baroclinic variables.

Finally, the truncated system (\ref{eq_trunc}) can be
expanded in powers of $\delta$, similar to the expansion
(\ref{asym_2d_1})--(\ref{asym_2d_3})
of the 2-D system (\ref{eq_long}).
The expansion of (\ref{eq_trunc})
is presented in Appendix~\ref{sec_trun_exp}
in (\ref{eq_merid_exp_1})--(\ref{eq_merid_exp_3}).
The next step in the method of multiscale asymptotics
is to solve the auxiliary problem that arises
in each of
(\ref{eq_merid_exp_1}), (\ref{eq_merid_exp_2}), and (\ref{eq_merid_exp_3}).

\section{Auxiliary problem: a degenerate system with constraint equation}\label{sec_aux_new}
In this section, the auxiliary problem is introduced.
It is a key part of the multi-scale analysis procedure where a forced linear system must be solved
(see chapter 5 of \cite{m03} for other examples).
In the linear system of the present paper,
a difficulty arises because of
a constraint equation; specifically, a constraint arises from 
meridional geostrophic balance where $\vi_{1t}$ does not appear 
in the leading order linear operator in (\ref{asym_2d_1}) or (\ref{eq_merid}).
This constraint equation presents a difficulty for directly computing the eigenmodes of the linear system.
To overcome this difficulty,
a reformulation of the auxiliary problem is presented here to 
enable direct calculation of the eigenmodes
and hence a direct solution of the forced linear system.

The linear system in $\vec{U}$ for the baroclinic component from (\ref{eq_trunc}) can be denoted as:
\begin{equation}
\NN \vec{U}_t + \LLU \vec{U} = \FFU.
\label{aux_abs_u}
\end{equation}
Notice the degenerate diagonal $8\times 8$ matrix 
$\NN = \text{diag}(1,1,1,1,0,1,1,1)$, which is degenerate due to the
$0$ in the $\vi'$ element of the diagonal.
Here $\FFU$ is the forcing term, 
including long-time dependency and bilinear terms.
Also recall from (\ref{eq_trunc}) that $\vec{U}$ is the vector of 
baroclinic variables, defined as
$\vec{U} = ( \lo,\lt,\ro,\rt,\vi', \qo,\qt, \ao)$.
Due to the degenerate matrix $\NN$,
the eigenmodes of this system cannot be found directly
using the standard procedure.

In order to circumvent the degenerate matrix $\NN$,
the auxiliary problem can be reformulated in terms of new variables:
$$\vec{\tilde{W}} = (K, R, Q, A, \vi, \chi, \lt, \qt),$$
where $K = \ro$, $R = \sqrt{2}\lo+2\rt$, $Q = \qo$, $A = \ao$, $\chi = \lo-\sqrt{2}\rt$.
This reversible linear transformation can be written as $\vec{\tilde{W}} = \AAA\vec{U}$.
Then the system for $\vec{\tilde{W}}$ is 
\begin{equation}
\AAA\NN\AAA^{-1} \partial_t \vec{\tilde{W}} + \AAA\LLU\AAA^{-1} \vec{\tilde{W}} = \AAA \FFU,
 \label{aux_abs_tilde}
\end{equation}
which can be further simplified to 
\begin{equation}
 \partial_t \vec{W} +\LLW \vec{W} = \FFW,
 \label{aux_abs}
\end{equation}
where $\vec{W} = (K, R, Q, A, \lt, \qt)$, 
and $\LLW$ is the spatial linear operator.
From $\vec{\tilde{W}}$ to $\vec{W}$, two variables $\vi$ and $\chi$ are eliminated because of the constraint equation.
For details, see Appendix~\ref{sec_aux}.

Note that this transformation of variables from $\vec{U}$ to $\vec{W}$
is used in the formulation of the MJO skeleton model \cite{ms09pnas,sm15},
and it has some similarities to the transformation of the
two-dimensional incompressible fluid flow equations from
velocity variables to vorticity/streamfunction variables
(see chapter 9 of \cite{m03}).
For the purposes here, the formulation in terms of 
$\vec{W} = (K, R, Q, A, \lt, \qt)$
allows easy identification of the linear eigenmodes of 
the linear operator $\LLW$,
since the $(K, R, Q, A)$ sub-system of (\ref{aux_abs})
is the linearized MJO skeleton model that has been
previously studied \cite{ms09pnas}.

Finally, the forcing $\FFU$ in (\ref{aux_abs_u})
must be transformed into $\FFW$ in (\ref{aux_abs}).
This transformation is needed in order to solve the
auxiliary problem at each order of magnitude in the
asymptotic expansion. 
See Appendix~\ref{sec_aux}
for the explicit formulas for $\FFW$
for each order of magnitude of the asymptotic expansion.

\section{Eigenmodes for the linear system} \label{sec_eig}
In this section, the eigenmodes are described for the baroclinic system and barotropic system.
(Later, in Section~\ref{sec_reduced}, these linear eigenmodes will be used in the identification of three-wave resonances.)

\begin{figure}
\centering
\includegraphics[width = .65\hsize]{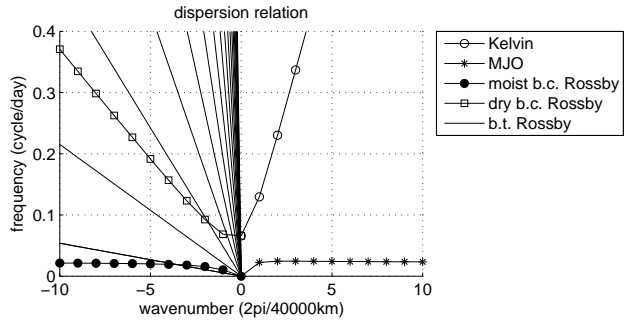}
\caption{Dispersion relation of the KRQA system and barotropic wind with different meridional wavelengths.}
\label{fig_disp}
\end{figure}

First, in the baroclinic system (\ref{aux_abs}) (with $\FFW= \vec{0}$),
the components of vector
$\vec{W} = (K, R, Q, A, \lt, \qt)$
can be separated into two groups:
$(K, R, Q, A)$ and $(\lt, \qt)$.
The KRQA system is closed
(as can be seen in its explicit formulation in Appendix~\ref{sec_aux}), 
and its four eigenmodes have been
described previously \cite{ms09pnas}:
dry Kelvin, MJO, moist Rossby and dry Rossby, 
 as shown in figure~\ref{fig_disp}.
In brief, the dry Kelvin wave is a fast eastward-propagating wave;
the MJO is a slow east-propagating wave; and the
moist and dry baroclinic Rossby waves are slow and fast
westward-propagating waves, respectively.
In addition to the KRQA system, $\lt$ satisfies 
an independent (decoupled) equation,
and $\qt$ is slaved to $(K, R, Q, A)$ and $\lt$
(see the explicit formulation in Appendix~\ref{sec_aux}).
Therefore the eigenmodes from the KRQA system of~\cite{ms09pnas} are not 
affected by the additional two variables
$\lt$ and $\qt$.
Notice that the baroclinic Rossby and Kelvin waves in this model are
dry equatorial waves, not
convectively coupled equatorial waves (CCEWs)~\cite{ketal09}.
The model here is a planetary-scale model that does not
explicitly resolve CCEWs, which occur mainly on smaller,
synoptic scales.
Nevertheless, one would expect similar wave interactions to hold for CCEWs,
due to the similar structures of dry and convectively coupled waves, 
if a model were used that
resolved such synoptic-scale convectively coupled waves.

Second, for the barotropic system (\ref{eq_trunc_barotropic}), 
the dispersion relation is written explicitly as
\begin{equation}
\obt(k) = -\frac{k}{L^2}.
\label{eq_obt}
\end{equation}
The above formula implies that
barotropic Rossby waves travel
faster for
smaller meridional wavenumber $L$ 
or, equivalently, longer meridional wavelength $Y$.
In figure~\ref{fig_disp},
barotropic dispersion relation are shown with different meridional wave numbers.

\section{The reduced asymptotic models}\label{sec_reduced}
In this section, the following reduced asymptotic model is derived
and its basic features are described:
\begin{subequations}
\begin{align}
&- \partial _{T_2}\beta {\color{white}+i d_1\beta_{j_11}\beta_{j_11}^{*2}}+ i d_2 \beta + i d_3 \alpha_{1}^* \alpha_{2}^*  = 0, \\
& -\partial _{T_2}\aji + i d_4 \aji^2 \alpha_{1}^*  + i d_5\aji + i d_6 \beta^*  \alpha_{2}^* = 0, \\
&-\partial _{T_2}\ajt + i d_7\ajt^2 \alpha_{2}^*  +i  d_8\ajt + i d_9  \beta^* \alpha_{1}^* = 0,
\end{align}
\label{eq_3wv_ode}
\end{subequations}
where the coefficients $d_j$ are all real numbers.
This is a system of ODEs for three-wave resonance,
where $\beta$ is the complex amplitude of the barotropic Rossby waves, and
$\alpha_{1}$ and $\alpha_{2}$ are the complex amplitudes
of two baroclinic waves.
The superscript $*$ stands for complex conjugate. 
For the applications of importance here, one of the $\alpha$s
will correspond to the MJO, and this ODE system describes
its interaction with the other waves, with an aim toward
MJO initiation and termination.

In brief, the terms in (\ref{eq_3wv_ode}) fall into 
three groups that are linked to different features in the full 2D model
(\ref{eq_full}):
the cubic terms are associated with the nonlinear $q$-$a$ interactions;
the linear terms come from dispersive terms;
the quadratic terms rise from the nonlinear baroclinic-barotropic interactions.
Here, in particular, the cubic self-interaction terms are a novel feature 
that are not typically found in three-wave resonance ODEs,
and they arise here from the effects of water vapor $q$ and 
convective activity $a$.

In the remainder of this section,
two basic properties of the system are described -- energy conservation
and a family of equilibrium points -- and the derivation is sketched,
with details of the derivation shown in Appendix~\ref{sec_multi}.

\subsection{Energy conservation}
The ODE system (\ref{eq_3wv_ode}) satisfies the following energy 
conservation principle:
\begin{equation}
\frac{\partial}{\partial T_2}(\beta\beta^* + \aji\aji^* + \ajt\ajt^*) =0.
\end{equation}
This conserved quantity is a consequence that all coefficients are pure imaginary, 
and in particular, that
\begin{equation}
d_3 +  d_6 + d_9 =0,
\label{eq_d369}
\end{equation}
which is a key component for energy conservation
 in three-wave interaction equations~\cite{m03}.

\subsection{Equilibium points in ODE system}
In addition to the trivial equilibrium point of (\ref{eq_3wv_ode})
where $\beta=\alpha_{1}=\alpha_{2}=0$,
a family of nontrivial equilibrium points also exists:
\begin{subequations}
  \begin{align}
|\aji|^2 &= \frac{d_2^{2}d_5d_7+d_2d_3d_6d_8}{d_3^2d_6d_9-d_2^{2}d_4d_7}\\
|\ajt|^2 &= \frac{d_2^{2}d_4d_8+d_2d_3d_9d_5}{d_3^2d_6d_9-d_2^{2}d_4d_7}\\
\beta&=-\frac{d_3}{d_2}\aji^*\ajt^*
\end{align}
\label{eq_fix_pt}
\end{subequations}
Two examples for the nontrivial equilibrium points are considered here:
\begin{subequations}
\begin{align}
& |\aji| \approx 0.24, \quad |\ajt| \approx 0.20, \quad |\beta|\approx 0.0033 \text{ for case MRB,} \\ 
& |\aji| \approx 2.4, \quad |\ajt| \approx 7.5, \quad |\beta|\approx 0.34 \text{ for case MKB.} 
\end{align}
\end{subequations}
The MRB and MKB are cases for different resonance triads. 
For MRB, the M stands for MJO, the R stands for equatorial Rossby wave,
and the B stands for barotropic Rossby wave;
and for MKB, the M and B are the same, and the K stands for
equatorial Kelvin wave.
These two cases are further discussed 
in Section~\ref{sec_3wv_num}
 (although the simulations in Section~\ref{sec_3wv_num}
are not set up to illustrate the dynamics near these nontrivial equilibrium points).

In a linear stability analysis (not shown),
the system~(\ref{sec_3wv_num}) is neutrally stable when 
linearized around the trivial equilibrium point of $\vec{0}$.
On the other hand, when linearized around the two examples
of the nontrivial equilibrium points,
the system is unstable.
It would be interesting to further explore these cases and their
physical interpretations in the future.

\subsection{Derivation}
The derivation of (\ref{eq_3wv_ode}) starts by  
assuming a leading order barotropic component of the form
\begin{equation}
B_1 = \frac{1}{\sqrt{E_{\text{T}}}}{\beta}(T_1,T_2) e^{i\thbt} + \text{C.C.},
\label{bt_ansatz}
\end{equation}
where $\thbt = \kbt x'+ \obt t'$, and $E_{\text{T}}$ is the barotropic energy unit
$$E_{\text{T}} = 2YL^2, $$
and leading order baroclinic variables of the form
\begin{align}
\vec{W}_1 &= \aji e^{i\thji} \vec{r}_{1} + \ajt e^{i\thjt} \vec{r}_{2}+ \text{C.C.},
\label{bc_ansatz}
\end{align}
where $\theta_{j} = k_j x' + \omega_j t' $, 
and $\vec{r}$ represents a right eigenvector of the linear system 
with components
$$\vec{r}_{j} = (\hat{K}_{1,j},\hat{R}_{1,j},\hat{Q}_{1,j},\hat{A}_{1, j},\hat{\lt}_{1,j},\hat{\qt}_{1,j}), \quad j = 1, 2. $$
Here the arbitrary complex phase factor of each eigenvector is chosen 
so that $\hat{K}_{1,j}$ is a real number,
and the eigenvectors $\vec{r}_j$ are normalized with respect to
the energy as
\begin{align}
&E_j = \vec{r}_j^{\dag} \mathcal{H} \vec{r}_j=1, \quad j = 1, 2,
\end{align}
where $\mathcal{H}$ is the Hessian matrix of the conserved energy for the linear baroclinic system.

Next, two auxiliary problems must be solved,
(\ref{krqa_2}) and (\ref{krqa_3}) of Appendix~\ref{sec_aux},
one for each of the two long time scales,
$T_1=\delta t'$ and $T_2=\delta^2 t'$.
The solution of the auxiliary problems is the key step in
the method of multiscale asymptotics in order to 
suppress secular growth and guarantee a consistent
asymptotic expansion of the variables~\cite{m03}.
As part of the second auxiliary problem,
the three waves must satisfy the following
resonance conditions:
\begin{subequations}
\begin{align}
\kji+\kjt+\kbt &=0, \\
 \oji+\ojt + \obt &=0.
 \end{align}
 \label{eq_3wv_ko}
\end{subequations}
Further details of derivation are presented in Appendix~\ref{sec_multi}.

\section{Validation study: the wavenumber--2 MJO mode}\label{sec_valid}

In this section, a special case is considered in order to
explore the time scales of validity of the ODE system in (\ref{eq_3wv_ode}).
If only a single baroclinic mode is considered, without a barotropic mode,
the ODE system in (\ref{eq_3wv_ode}) becomes a single ODE:
\begin{equation}
 -\partial _{T_2}\aji + i d_4 \aji^2\aji^* + i d_5\aji = 0.
\label{eq_3wv_ode_ms11}
\end{equation}
The energy conservation of this single mode then takes the form
\begin{equation}
\frac{\partial}{\partial T_2}(\aji\aji^* ) =0.
\end{equation}
This special case can then be compared against 
numerical solutions that have been presented previously elsewhere~\cite{ms11},
which will be regarded as the `true' solution here.
The specific case chosen is Case U2 of~\cite{ms11},
which has a wavenumber-2 MJO eigenmode as the initial condition
for a nonlinear simulation.

A few technical details for consistency with~\cite{ms11} are the following.
The coefficient $c_1$ in (\ref{eq_merid}) is taken to be $1$ 
for consistency with the 
equation $A_t=\Gamma Q A$ used
in~\cite{ms11}.
Also, since the model of \cite{ms11} does not include the dispersive
effects of $\vi$ (although it does include other dispersive effects), 
the coefficient $d_5$ in the ODE system (\ref{eq_3wv_ode_ms11})
is set to 0.
As in \cite{ms11}, here the computational domain is $[0, X]$, 
where $X = 40,000$~km in the circumference of Earth at the equator.

\begin{figure}[h]
\centering
\includegraphics[width = .4\hsize]{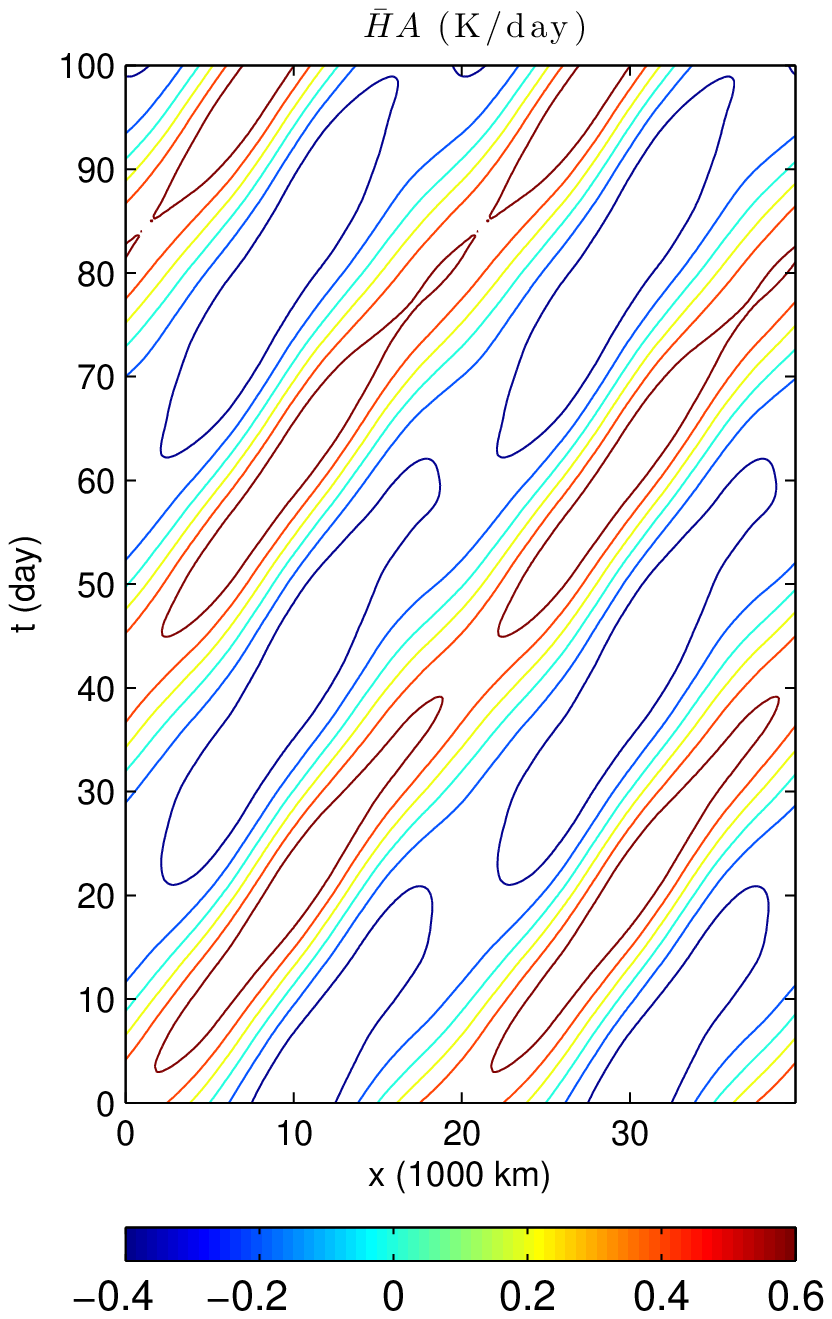}
\includegraphics[width = .4\hsize]{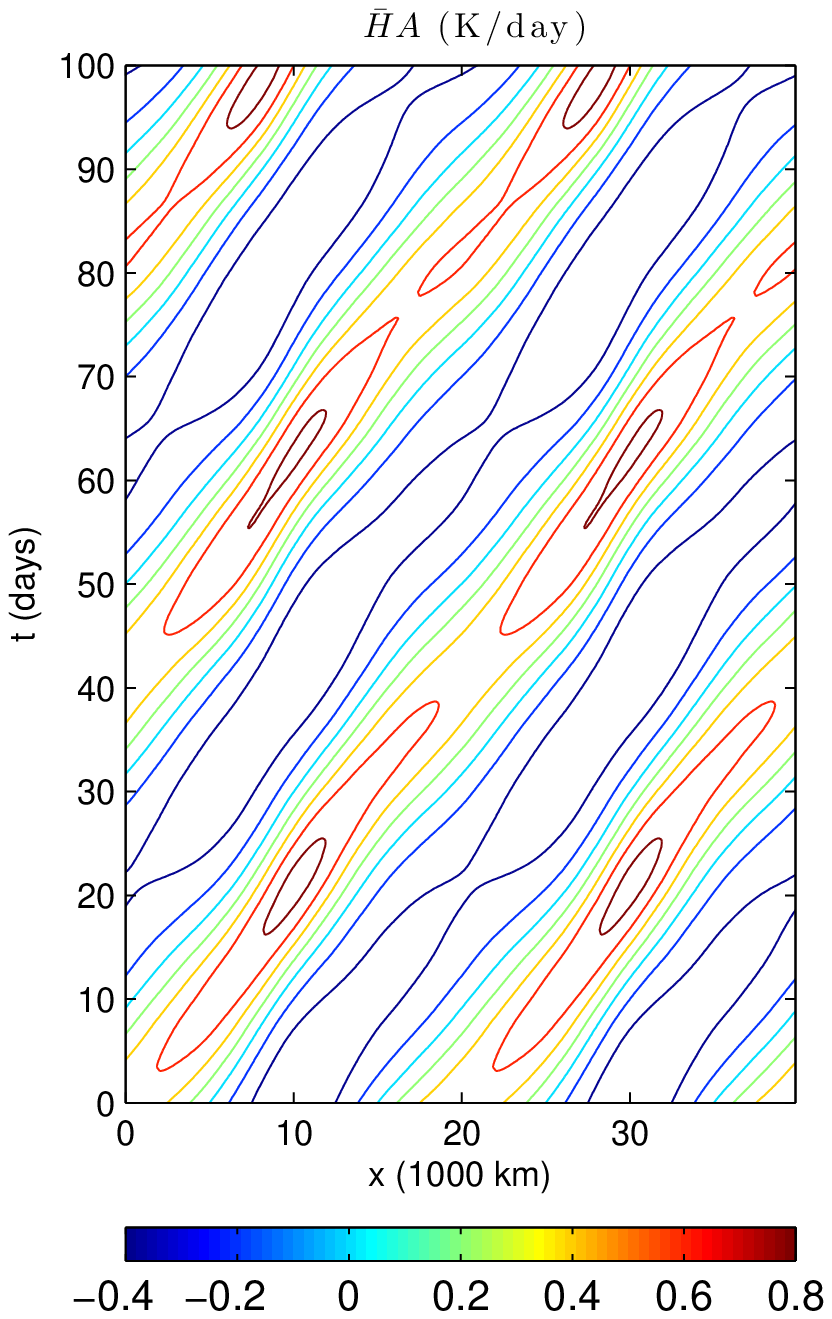}\\
\includegraphics[width = .4\hsize]{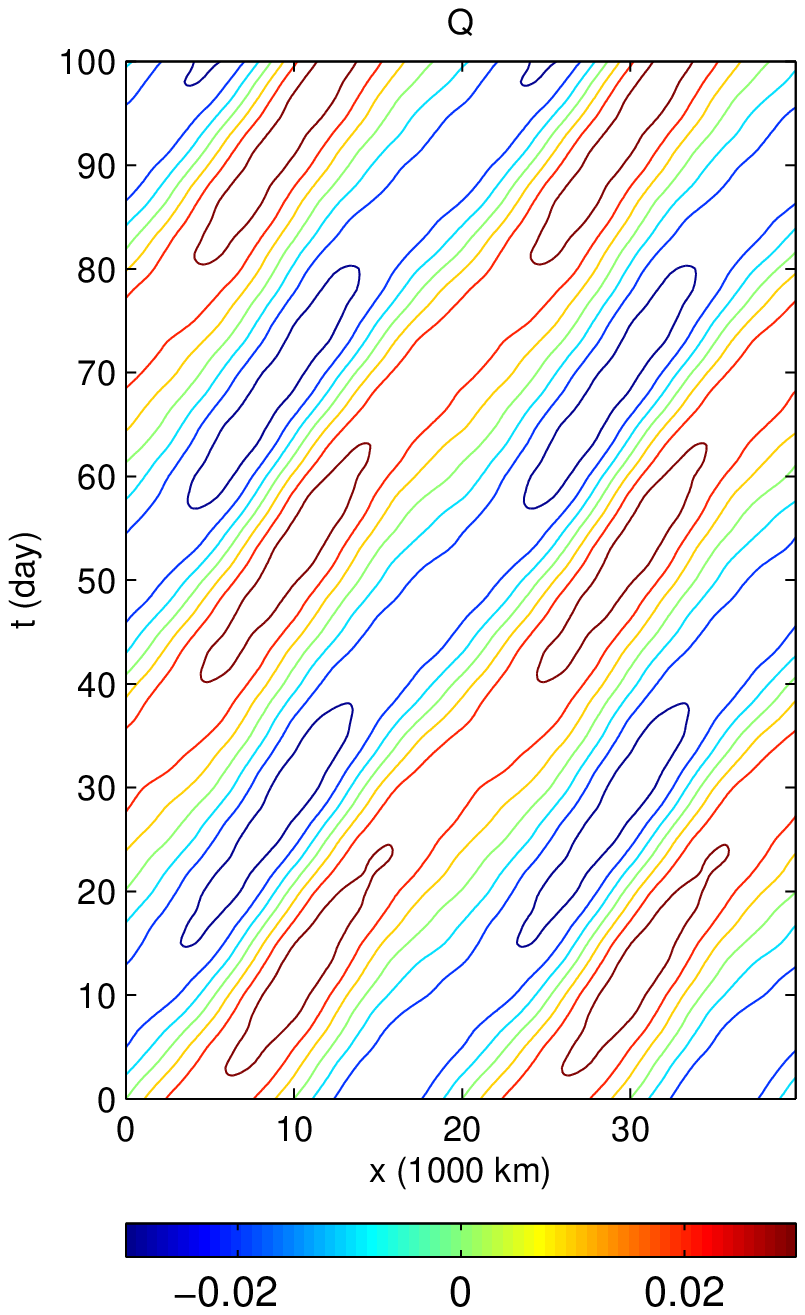}
\includegraphics[width = .4\hsize]{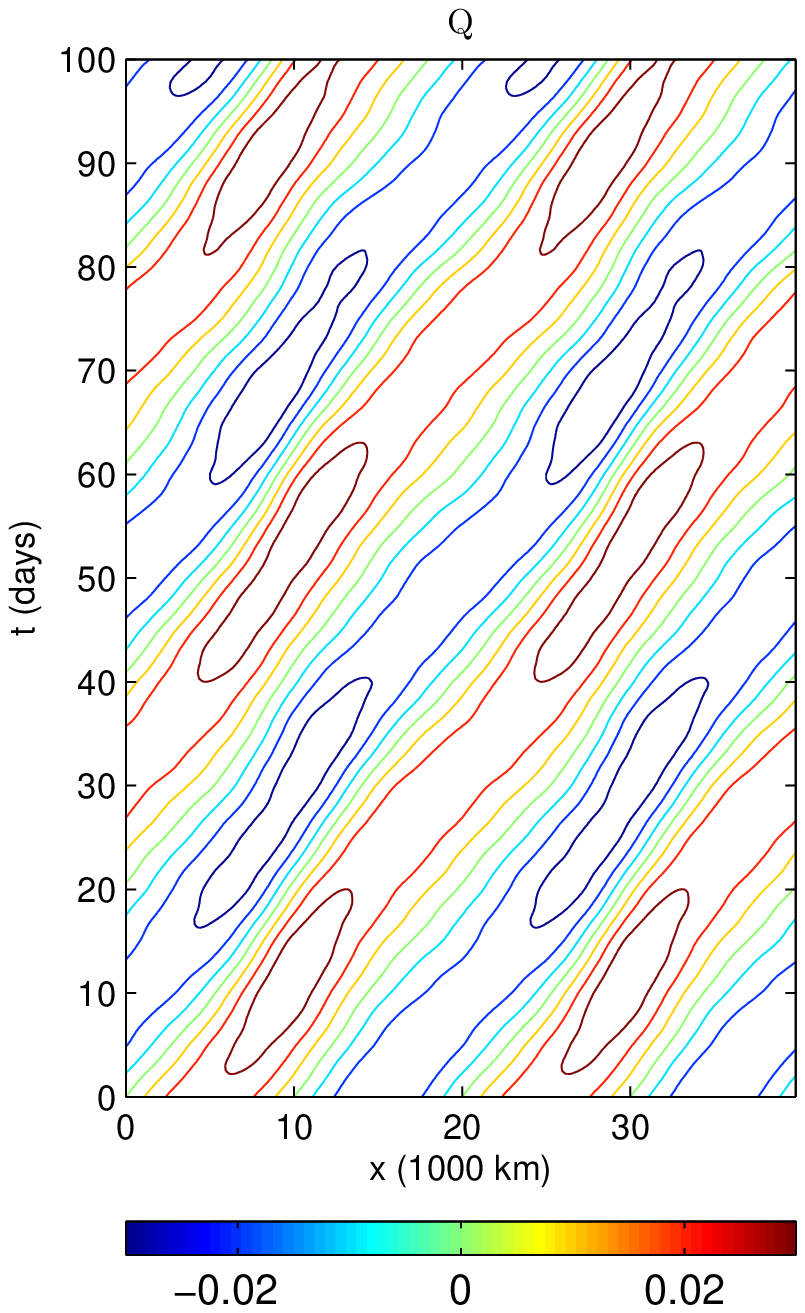}
\caption{Comparisons between asymptotic solutions (left) 
and numerical (`true') solutions (right) for $\Abar = 0.1$
for the case of a wavenumber-2 MJO mode.
The convective activity $\bar{H}A$ (top) and the water vapor $Q$ (bottom)
are shown as functions of $x$ and $t$.} 
\label{fig_mjo_k2_comp_A0b_1}
\end{figure}

\begin{figure}[h]
\centering
\includegraphics[width = .4\hsize]{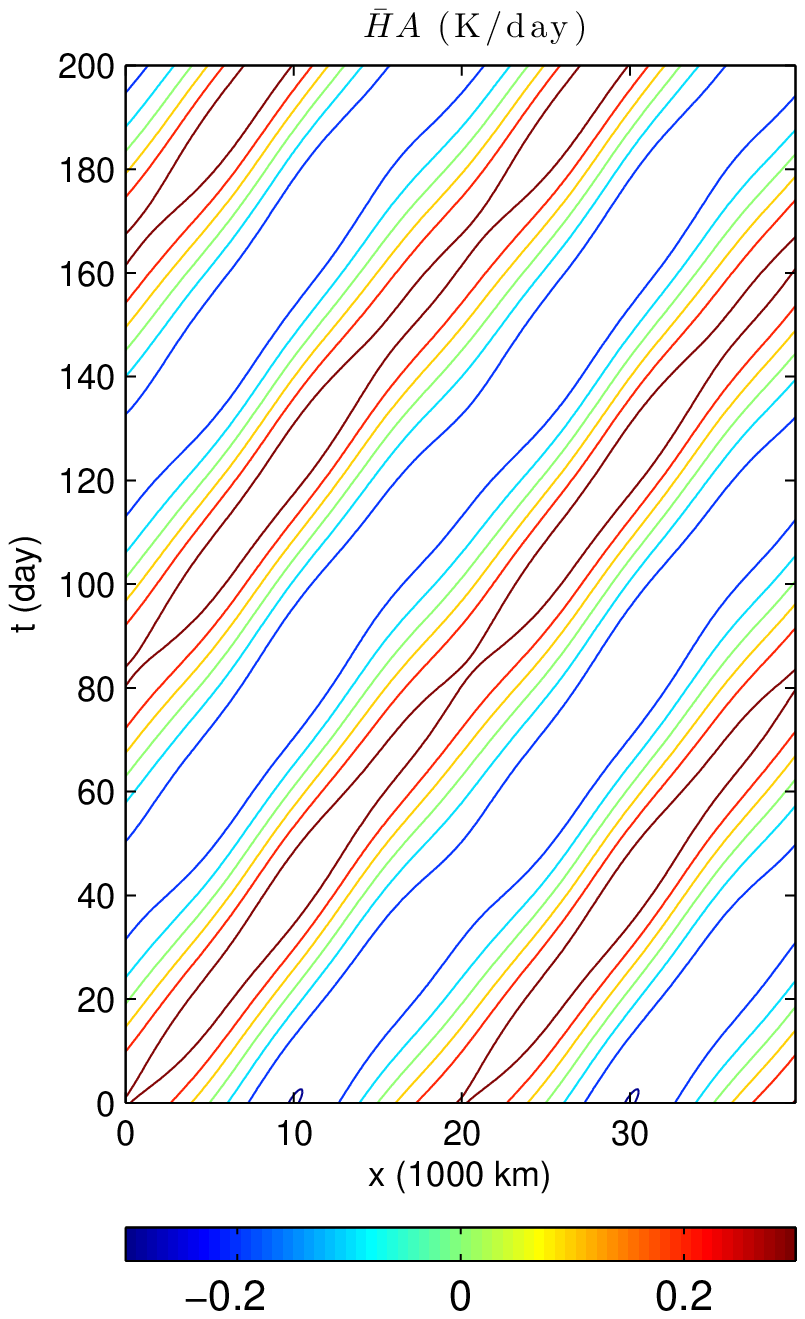}
\includegraphics[width = .4\hsize]{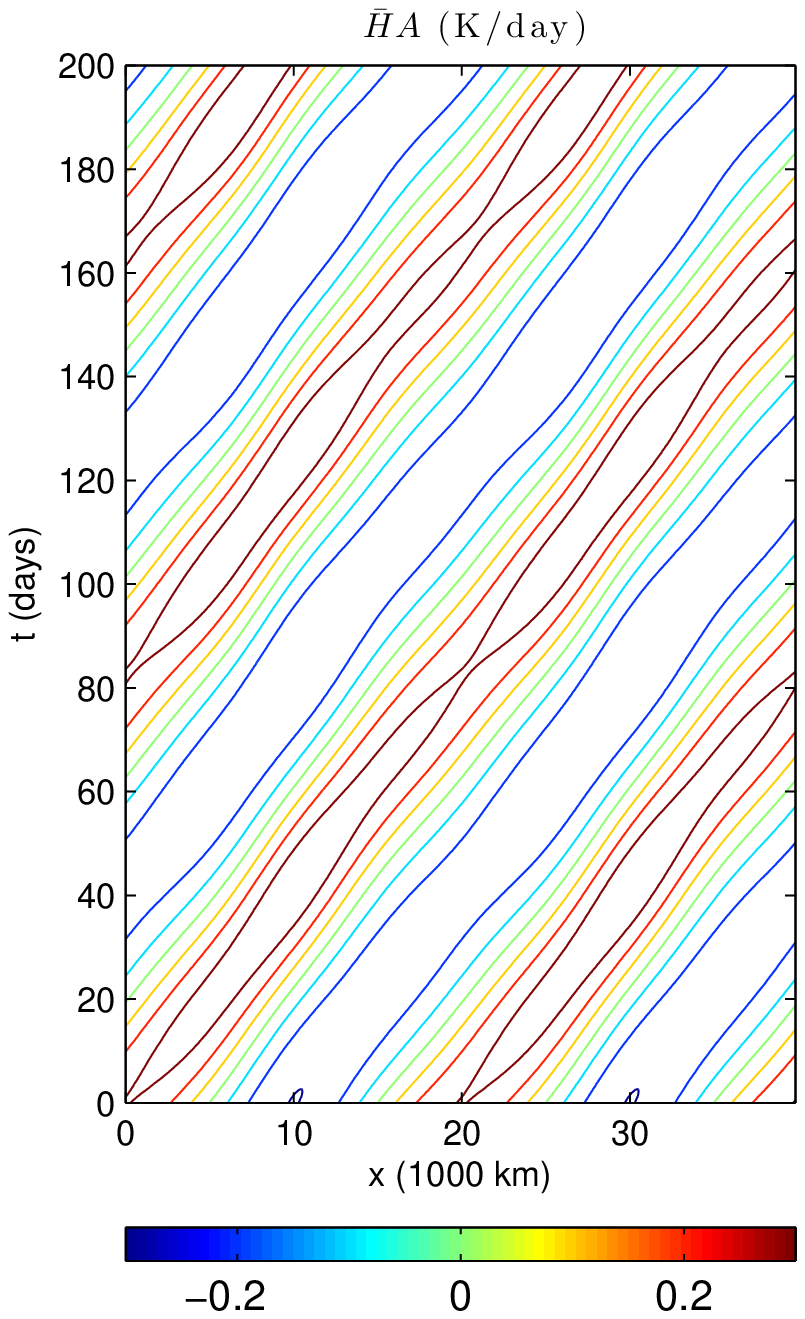}\\
\includegraphics[width = .4\hsize]{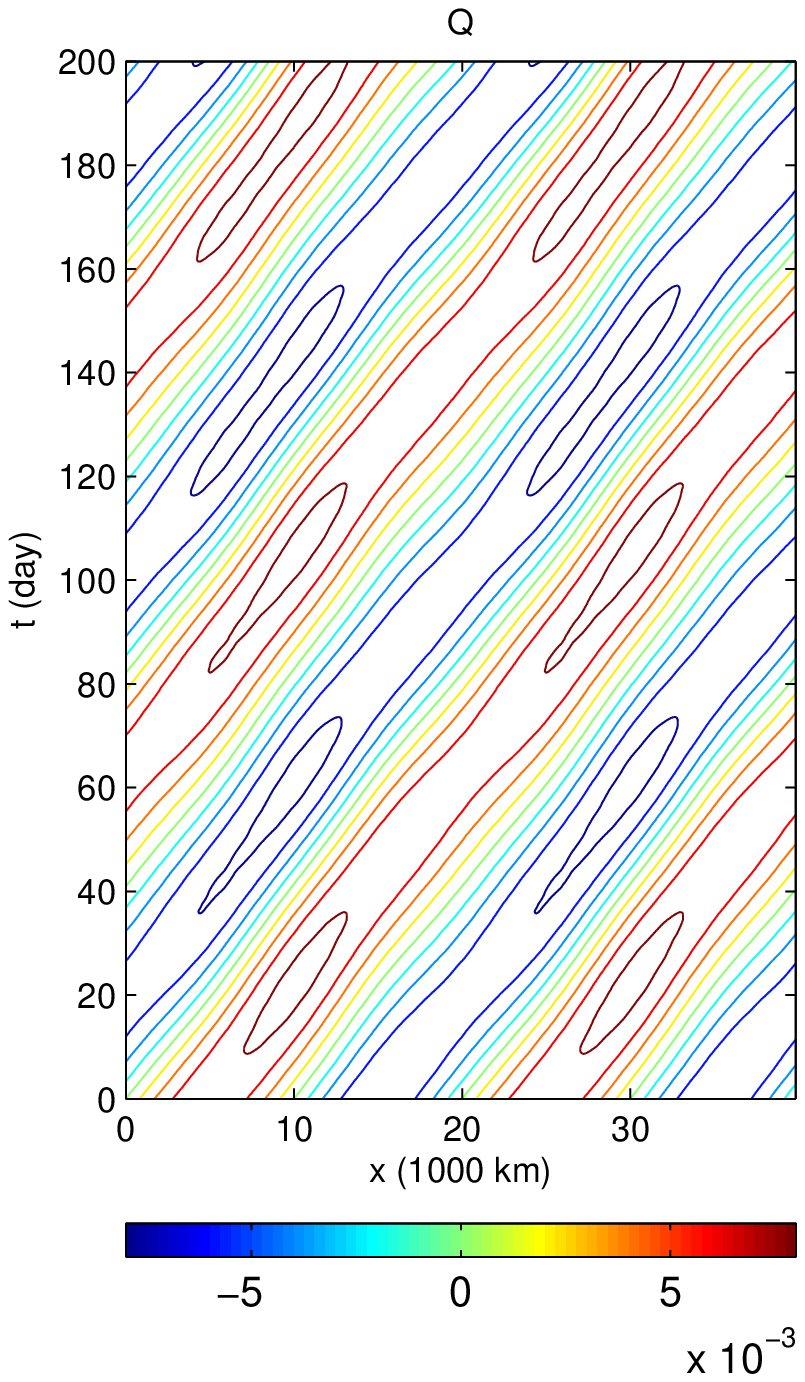}
\includegraphics[width = .4\hsize]{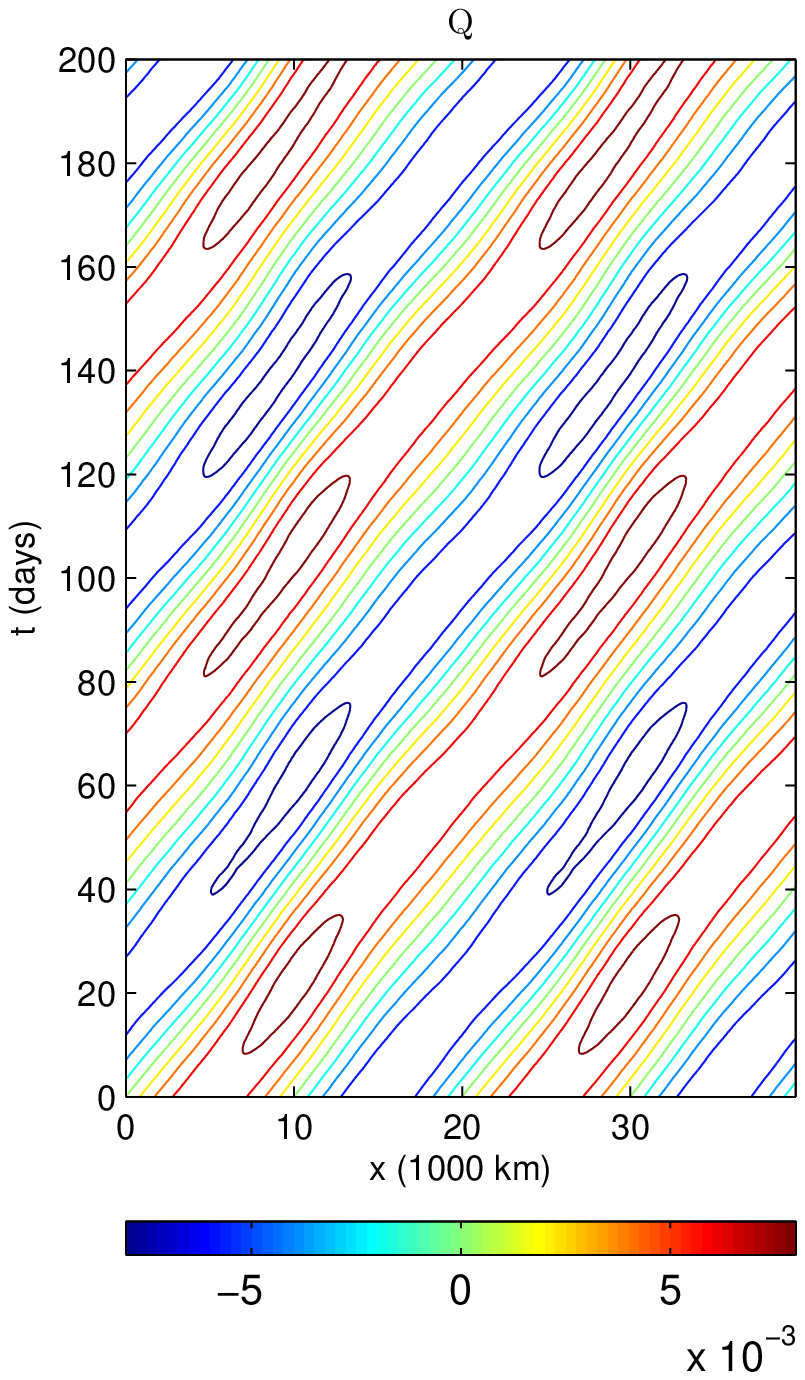}
\caption{Comparisons between asymptotic solutions (left) 
and numerical (`true') solutions (right) for $\Abar = 0.025$
for the case of a wavenumber-2 MJO mode.
The convective activity $\bar{H}A$ (top) and the water vapor $Q$ (bottom)
are shown as functions of $x$ and $t$.} 
\label{fig_mjo_k2_comp_A0b_025}
\end{figure}

For the validation study,
three different values of $\delta$ are chosen: 
$\delta=\sqrt{0.1}$, $\sqrt{0.1}/2$ and $\sqrt{0.1}/4$,
corresponding to three different RCE states: 
$\delta^2\abar=$ 0.1, 0.025 and 0.00625.
The solutions are evolved out to times 100, 200, and 400 days respectively, 
matching the long wave assumption.
Table~\ref{tb_compare_u2}
shows the relative error between the asymptotic solution and
the `true' solution.
The \textit{relative} error decreases by a factor of 2
as $\delta$ is decreased by a factor of 2,
indicating a \textit{relative} error of $O(\delta)$,
suggesting an \textit{absolute} error of $O(\delta^3)$ 
and a true solution with magnitude of $O(\delta^2)$.
Figures~\ref{fig_mjo_k2_comp_A0b_1} and \ref{fig_mjo_k2_comp_A0b_025} 
show comparisons between the asymptotic solutions and 
numerical (`true') solutions for
$\delta=\sqrt{0.1}$ and $\sqrt{0.1}/2$
(with $\delta=\sqrt{0.1}/4$ not shown since it is indistinguishable
from the `true' solution).
In Figure~\ref{fig_mjo_k2_comp_A0b_1},
the agreement is good out to 100 days,
although some of the peaks in convective activity $\bar{H}A$
are not captured by the asymptotic solution.
In Figure~\ref{fig_mjo_k2_comp_A0b_025},
the agreement is excellent out to 200 days.

As a further validity test of the asymptotic model,
the same comparisons have been repeated (not shown) with
the addition of the second order corrections.
More specifically, whereas the asymptotic solution $\vec{W}_\text{asym}$
in Table~\ref{tb_compare_u2} was defined as
$\vec{W}_\text{asym}=\delta^2\vec{W}_1$,
the comparisons have been repeated using
$\vec{W}_\text{asym}=\delta^2\vec{W}_1+\delta^3\vec{W}_2$,
where the second order correction $\vec{W}_2$ is described
in more detail in Appendix~\ref{sec_aux} and \ref{sec_multi}.
In these new tests, the values of Table \ref{tb_compare_u2} change to
0.2685, 0.0627, 0.0183.
These values show a decrease in \textit{relative} error of a factor of 4
as $\delta$ is decreased by a factor of 2,
indicating a \textit{relative} error of $O(\delta^2)$,
in line with an \textit{absolute} error of $O(\delta^4$) 
and a true solution with magnitude of $O(\delta^2)$.

In summary, the asymptotic solutions have significant accuracy
on time scales of 100 days or longer,
which are roughly the time scales for application to
tropical--extratropical interactions and 
MJO initiation and termination.

\begin{table}
\centering
  \begin{tabular}{c|c|c|c}
% after \\ : \hline or \cline{col1-col2} \cline{col3-col4} ...
   & $\delta^2\abar=$0.1 at $100$~day & $\delta^2\abar=$0.025 at 200~day & $\delta^2\abar=$0.00625 at 400~day  \\
   \hline
 $\|\vec{W}_{\text{asym}}-\vec{W}_{\text{n}}\|/\|\vec{W}_{\text{n}}\|$   &   0.3890 & 0.1657 &0.0770 \\
% \hline
\end{tabular}
  \caption{The relative difference between numerical solution $\vec{W}_n$
(regarded as the `true' solution here) and the asymptotic solution
  ($\vec{W}_{\text{asym}}$)
  whose amplitude $\alpha_{j_1}(T_2)$ is governed by
  the ODE (\ref{eq_3wv_ode_ms11}).
  Three different values of $\delta$ are considered along with
  three different corresponding times.}
  \label{tb_compare_u2}
\end{table}

%\begin{table}
%\centering
%  \begin{tabular}{c|c|c|c}
%% after \\ : \hline or \cline{col1-col2} \cline{col3-col4} ...
%   & $\delta^2\abar=$0.1 at $100$~day & $\delta^2\abar=$0.025 at 200~day & $\delta^2\abar=$0.00625 at 400~day  \\
%   \hline
%$\|\vec{W}_{1}-\vec{W}_{\text{n}}\|/\|\vec{W}_{\text{n}}\|$   &   0.4008 & 0.1704 &0.0774 \\
% $\|\vec{W}_{2}-\vec{W}_{\text{n}}\|/\|\vec{W}_{\text{n}}\|$  &   0.2685 & 0.0773 &0.0181 \\
% $\|\vec{W}_{1\alpha}-\vec{W}_{\text{n}}\|/\|\vec{W}_{\text{n}}\|$   &   0.3490 & 0.1572 &0.0757 \\
%$\|\vec{W}_{2\alpha}-\vec{W}_{\text{n}}\|/\|\vec{W}_{\text{n}}\|$  &   0.1209 & 0.0329 &0.0076\\
%% \hline
%\end{tabular}
%  \caption{The difference of numerical solutions and asymptotic solutions at different times.
%  The numerical solution $\vec{W}_{\text{n}}$ has relative error within 4e-4 by varying CFL number.
%  Three asymptotic solutions are compared: 
%    linear solutions without $\alpha$ ($\vec{W}_{1}$), 
%  and traveling wave solution with second order correction without $\alpha$ ($\vec{W}_{2}$).
%  linear solutions with $\alpha$ ($\vec{W}_{1\alpha}$), 
%  and traveling wave solution with second order correction with $\alpha$ ($\vec{W}_{2\alpha}$).
%  The asymptotic behavior is by varying the amplitude of $\Abar$. }
%  \label{tb_compare_u2}
%\end{table}

\section{Three--wave interactions} \label{sec_3wv_num}
\begin{figure}
\begin{center}
\includegraphics[width = .45\hsize]{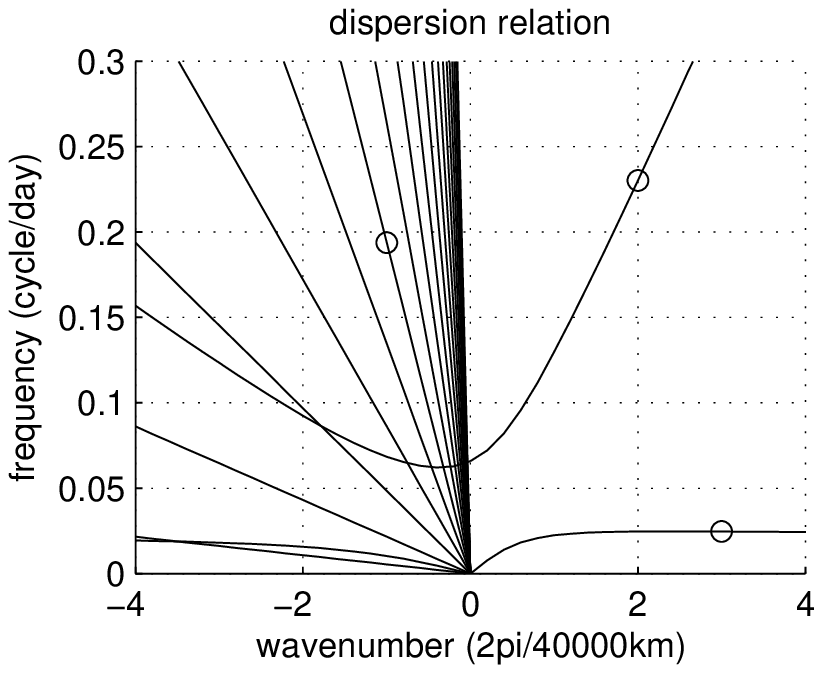}
\includegraphics[width = .45\hsize]{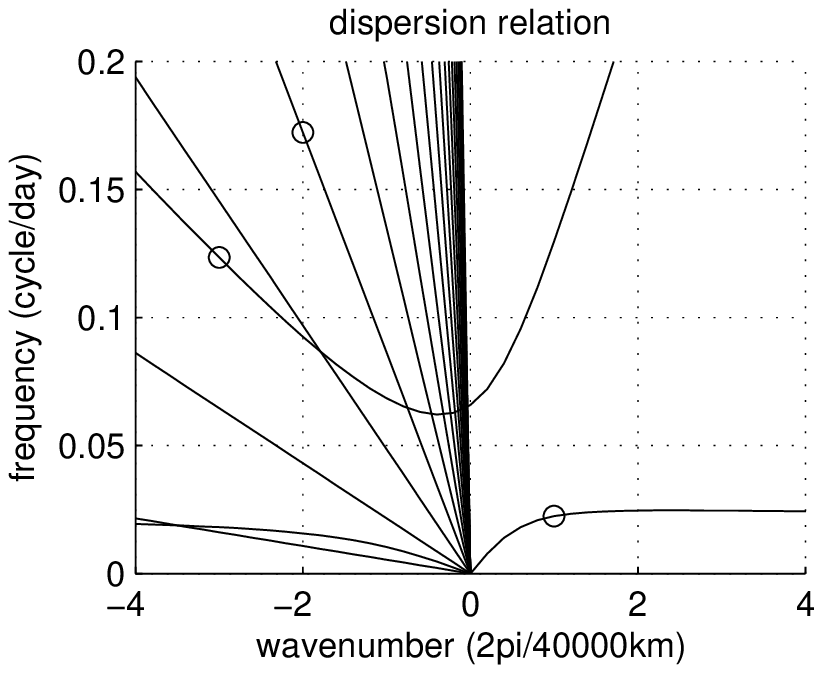}
\caption
{Resonance condition for three-wave interactions,
where circles indicate the particular wave numbers and frequencies
that leads to resonances.
Left: MJO, Kelvin and barotropic Rossby wave;
right: MJO, Rossby and barotropic Rossby wave.
}
\label{fig_disp_3wv}
\end{center}
\end{figure}

Here two types of three-wave interactions are considered as shown in figure~\ref{fig_disp_3wv}:
\begin{equation}
\begin{array}{ll}
\text{(1):}&\text{ MJO--dry baroclinic Rossby--barotropic Rossby waves (MRB)}\\
 &\omjo(1)+ \odr(-3)+\obt_{,Y\approx\text{2,000~km}}(2)=0\\
 &\\
 \text{(2):} &\text{ MJO--dry Kelvin--barotropic Rossby waves (MKB)}  \\
 & \omjo(-3)+\okk(2)+\obt_{,Y\approx \text{3,000~km}}(1)= 0 
\end{array}
\label{eq_3wv_12}
\end{equation}
For the acronym MRB, 
the M stands for MJO, the R stands for equatorial Rossby wave,
and the B stands for barotropic Rossby wave;
and for MKB, the M and B are the same, and the K stands for
equatorial Kelvin wave.
The meridional wavelength $2Y$ of the barotropic Rossby wave is 
chosen so that the resonance conditions (\ref{eq_3wv_ko})
are exactly satisfied.
The numerical results here are focused on the ODE system (\ref{eq_3wv_ode}),
with wave amplitudes $\beta$, $\aji$, and $\ajt$,
where the energy exchange between three waves is of the most interest.
The coefficients in (\ref{eq_3wv_ode}) for the two cases are in table~\ref{tb_coef3wave}.
The RCE $\abar$ is chosen to be $\abar = 0.1$ so that the small parameter $\delta = \sqrt{0.1}$.

\begin{table}
\begin{tabu}{c|cccccccc}
%\tabucline[1pt]{-}
&  $d_2$ & $d_3$& $d_4$&$d_5$&$d_6$&$d_7$&$d_8$&$d_9$\\
 \hline
MRB & $-0.45$ & $-8.3e$-$3$ & $-5.2e$-$2$ & $1.7e$-$3$ & $-0.28$ & $-2.4e$-$2$ & $1.37$ & $0.29$ \\
 MKB & $0.45$ & $-3.2e$-$2$ & $-0.11$& $6.0e$-$3$ & $0.11$ & $7.8e$-$3$ & $-8.5e$-$7$ & $-7.6e$-$2$\\
 %\tabucline[1pt]{-}
\end{tabu}
\caption{Coefficients $d_j$ for the 3-wave interaction ODEs (\ref{eq_3wv_ode})
for two cases: 
the MJO, dry baroclinic Rossby, and barotropic Rossby wave interactions (MRB)
(top row)
and the MJO, dry Kelvin, and barotropic Rossby wave interactions (MKB)
(bottom row).}
\label{tb_coef3wave}
\end{table}

\subsection{MJO--dry baroclinic Rossby--barotropic Rossby waves (MRB)} \label{sec_MRB}
\begin{figure}
\begin{center}
\includegraphics[width = .45\hsize]{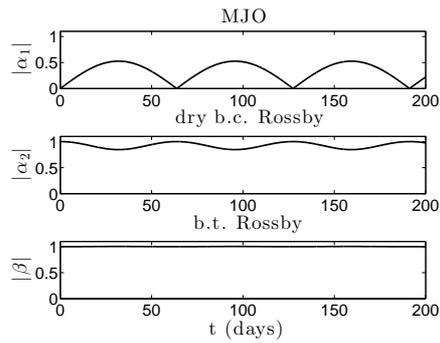}
\caption{MJO--Rossby--barotropic Rossby wave interactions.
Initial condition: $\aji$ = 0, $|\ajt_{(t=0)}| =|\beta_{(t=0)}|=1$.}
\label{fig_MRB_r1}
\end{center}
\end{figure}

In this case, $\aji$ and $\ajt$ are the amplitudes for the MJO 
and dry baroclinic Rossby waves, respectively.
To consider MJO initiation, $\aji$ is set to be zero for the 
initial condition in the ODE,
and the other waves have amplitudes of $|\ajt_{(t=0)}|=|\beta_{(t=0)}|=1 $.
The results of the numerical simulation are shown in
Figure~\ref{fig_MRB_r1}.
The waves exchange energy periodically, with period of about~60~days,
roughly the time scale for MJO initiation and termination in nature.
At early times, the MJO amplitude grows (i.e., the MJO initiates) 
by extracting energy from the
other two waves, mainly from the dry baroclinic Rossby wave.
Termination of the MJO then follows from a transfer of energy
back to the dry baroclinic Rossby wave.

\begin{figure}
\begin{center}
\includegraphics[width = .45\hsize]{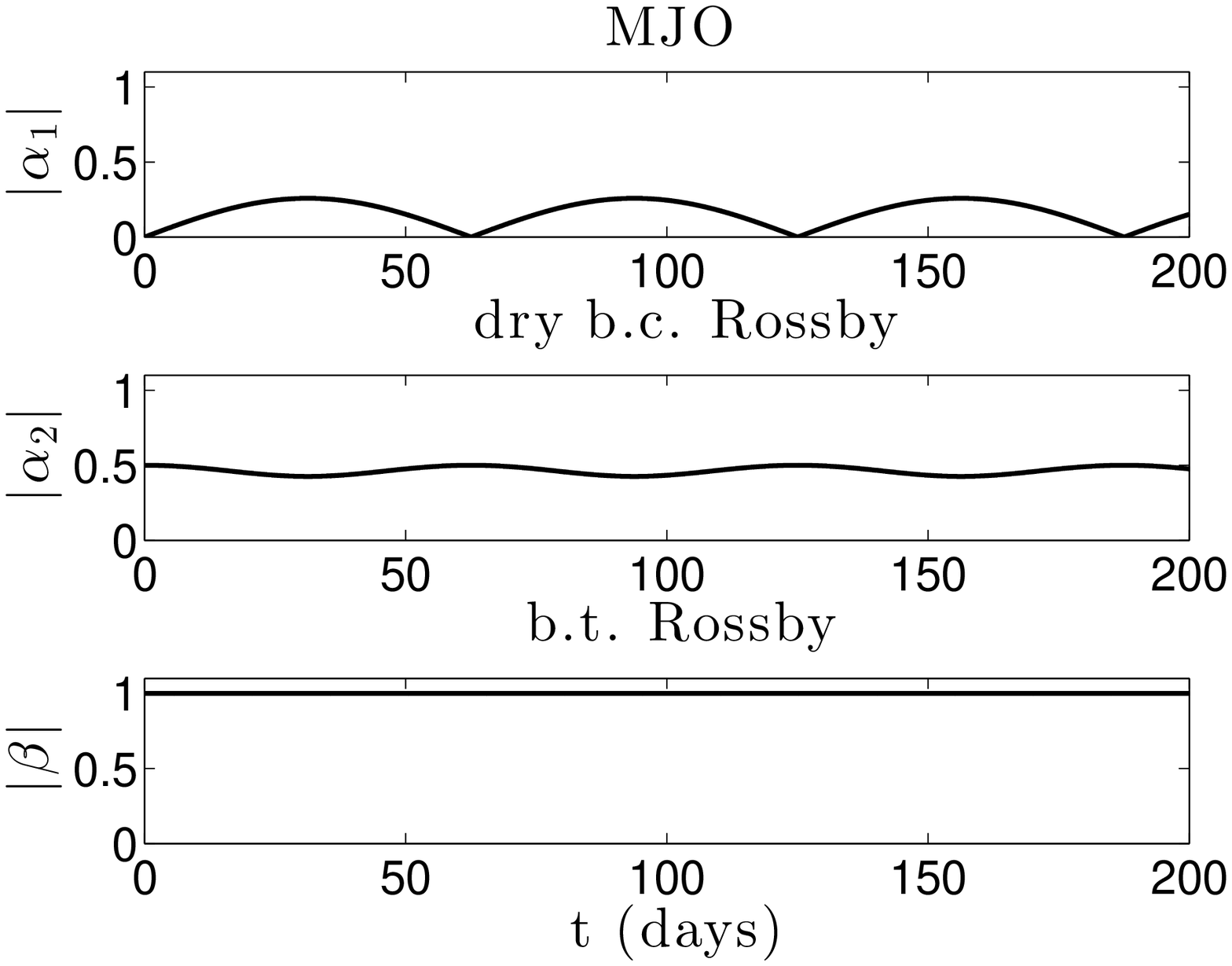}
\includegraphics[width = .45\hsize]{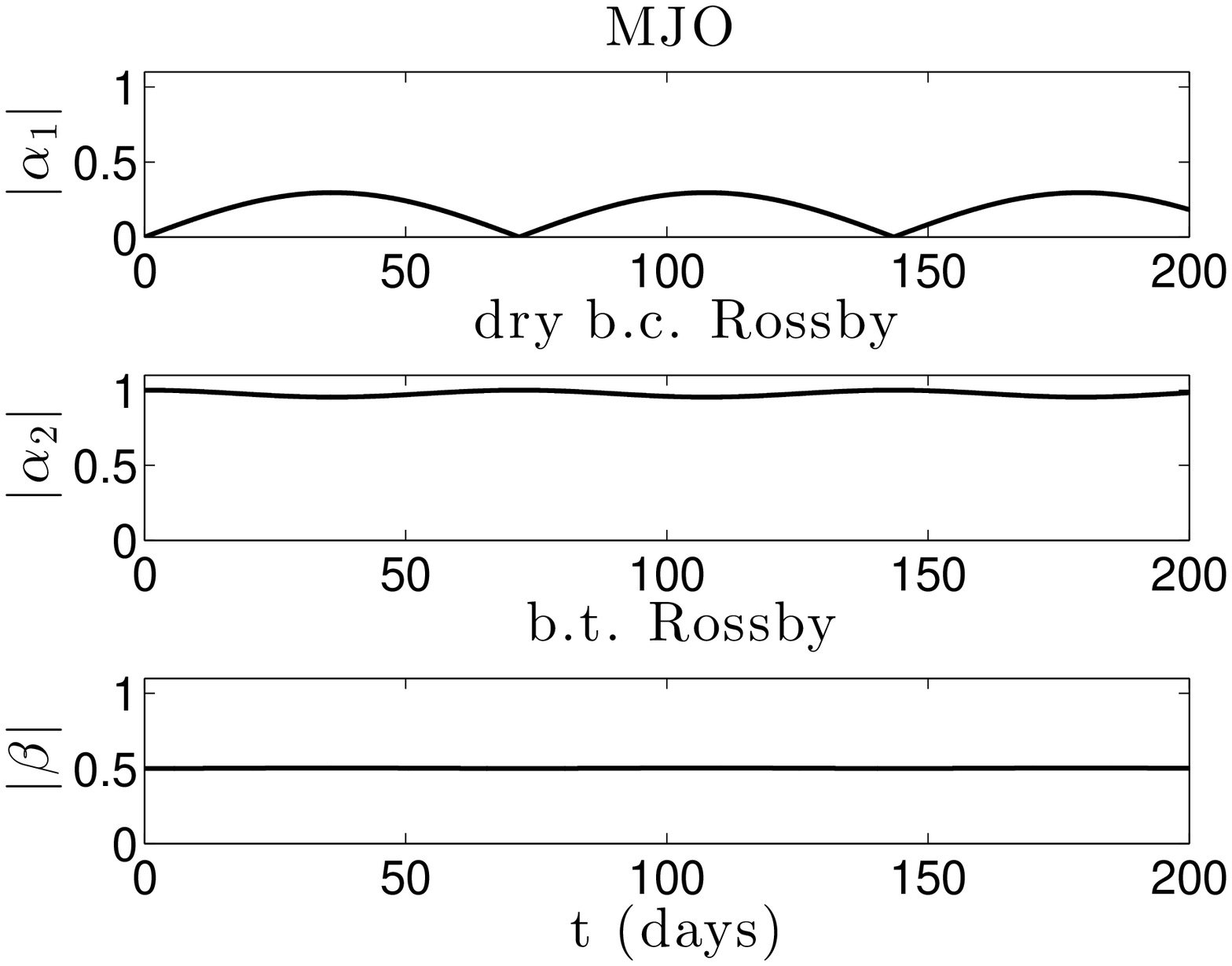}
\hrule
\includegraphics[width = .45\hsize]{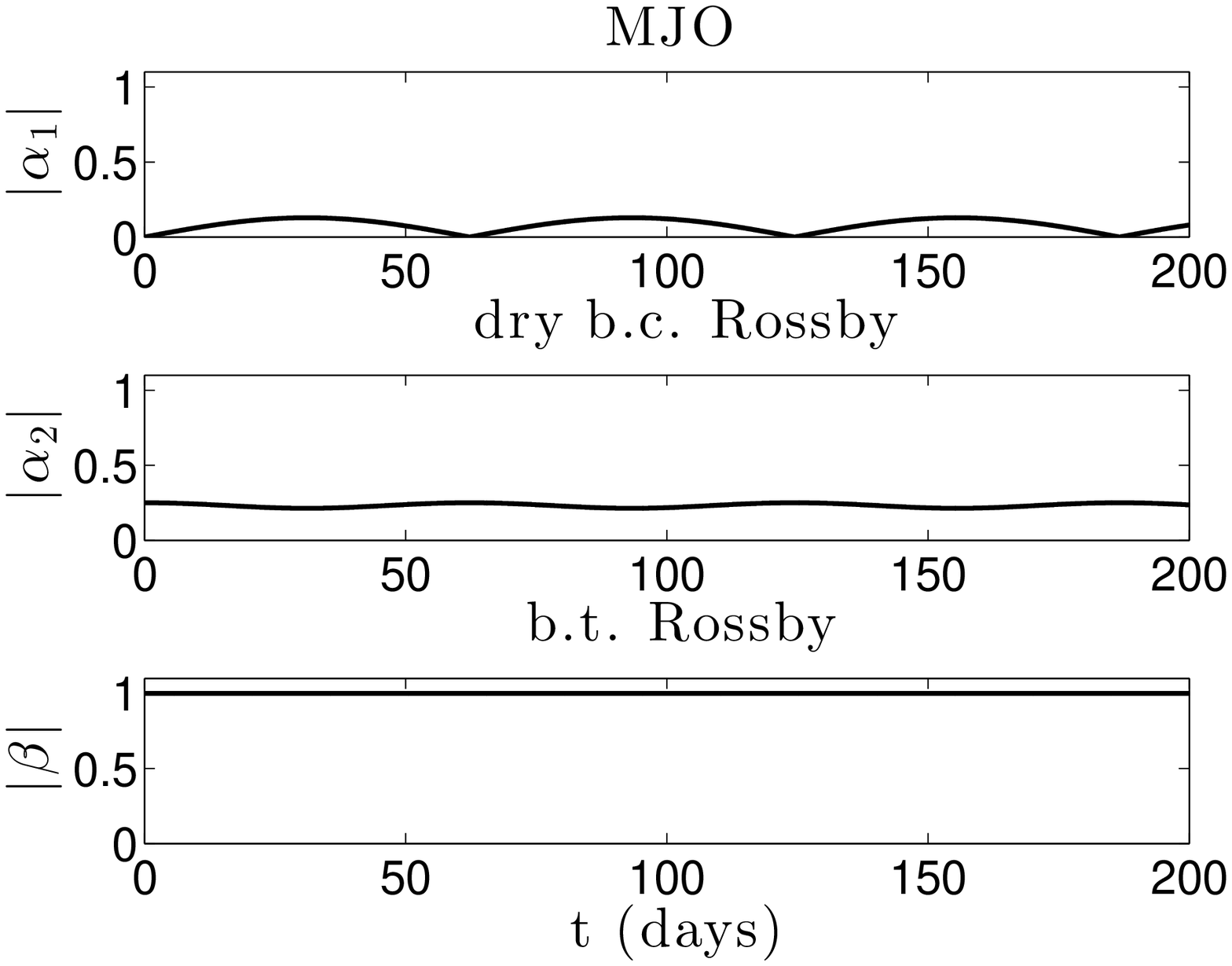}
\includegraphics[width = .45\hsize]{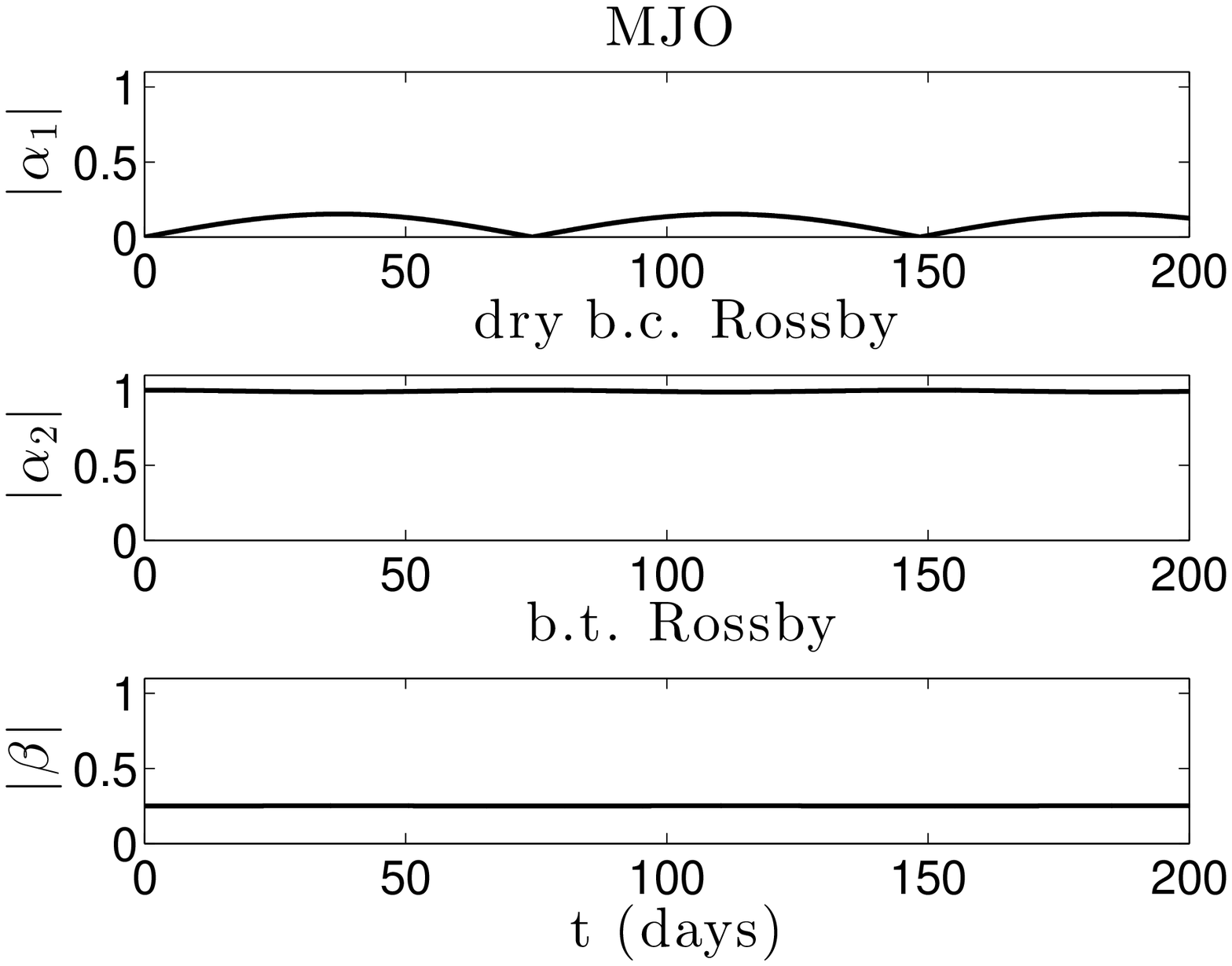}
\caption
{MJO--Rossby--barotropic Rossby wave interactions.
Initial condition: $\aji$ = 0.
 (a) $|\ajt_{(t=0)}| =0.5$,$ |\beta_{(t=0)}|= 1$;
 (b) $|\ajt_{(t=0)}| =1$, $|\beta_{(t=0)}|= 0.5$;
 (c) $|\ajt_{(t=0)}| =0.25$, $|\beta_{(t=0)}|= 1$;
 (d) $|\ajt_{(t=0)}| =1$, $|\beta_{(t=0)}|= 0.25$.
}
\label{fig_MRB_rdiff}
\end{center}
\end{figure}

To explore the sensitivity of these results, 
different initial conditions are chosen, 
and the results are in figure~\ref{fig_MRB_rdiff}.
The initial amplitudes for baroclinic Rossby wave 
and barotropic Rossby waves are
 (a) $|\ajt_{(t=0)}| =0.5$,$ |\beta_{(t=0)}|= 1$;
 (b) $|\ajt_{(t=0)}| =1$, $|\beta_{(t=0)}|= 0.5$;
 (c) $|\ajt_{(t=0)}| =0.25$, $|\beta_{(t=0)}|= 1$;
 (d) $|\ajt_{(t=0)}| =1$, $|\beta_{(t=0)}|= 0.25$.
The simulations are performed with different initial phases 
of $\ajt$ and $\beta$, but this has no effect on 
the evolution of the amplitudes.
In the simulations in Figure \ref{fig_MRB_rdiff}, 
it can be seen that the amount of energy 
extracted by the MJO varies roughly proportionally to
the amount of energy present initially in the dry baroclinic Rossby wave
and barotropic Rossby wave.
On the other hand, the time scale for the energy exchange
changes very little among these cases.
In all cases, the energy of the barotropic Rossby wave
stays essentially constant and is not transferred to the MJO.
These results indicate that the eventual strength of the MJO
depends on the amplitude of the barotropic Rossby waves,
but the MJO acquires its energy from the baroclinic rather than
barotropic Rossby waves.

\subsection{MJO--dry Kelvin--barotropic Rossby waves (MKB)}
\begin{figure}
\begin{center}
\includegraphics[width = .45\hsize]{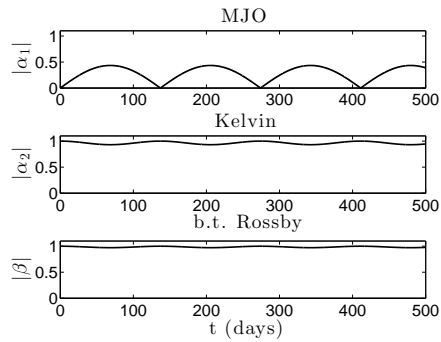}
\caption{MJO--Kelvin--barotropic Rossby wave interactions.
Initial condition: $\aji$ = 0, $|\ajt_{(t=0)}| =|\beta_{(t=0)}|=1$.}
\label{fig_MKB_r1}
\end{center}
\end{figure}

In this case, $\aji$ and $\ajt$ are the amplitudes for 
the MJO and dry baroclinic Kelvin waves, respectively.
Again considering MJO initiation,
the initial conditions are $\aji=0$ and
$|\ajt_{(t=0)}| =|\beta_{(t=0)}|=1 $, and the results are shown in
Figure~\ref{fig_MKB_r1}.
While the solution is similar in character to the MRB case,
two main differences can be seen.
First, the temporal period of roughly 120 days is twice that of
the MRB case but still within the range of time scales
for MJO initiation and termination in nature.
Secondly, both the dry Kelvin wave and the barotropic Rossby waves 
contribute to the growth of MJO, 
although the energy exchange with the barotropic Rossby wave
is somewhat small.

\begin{figure}
\begin{center}
\includegraphics[width = .45\hsize]{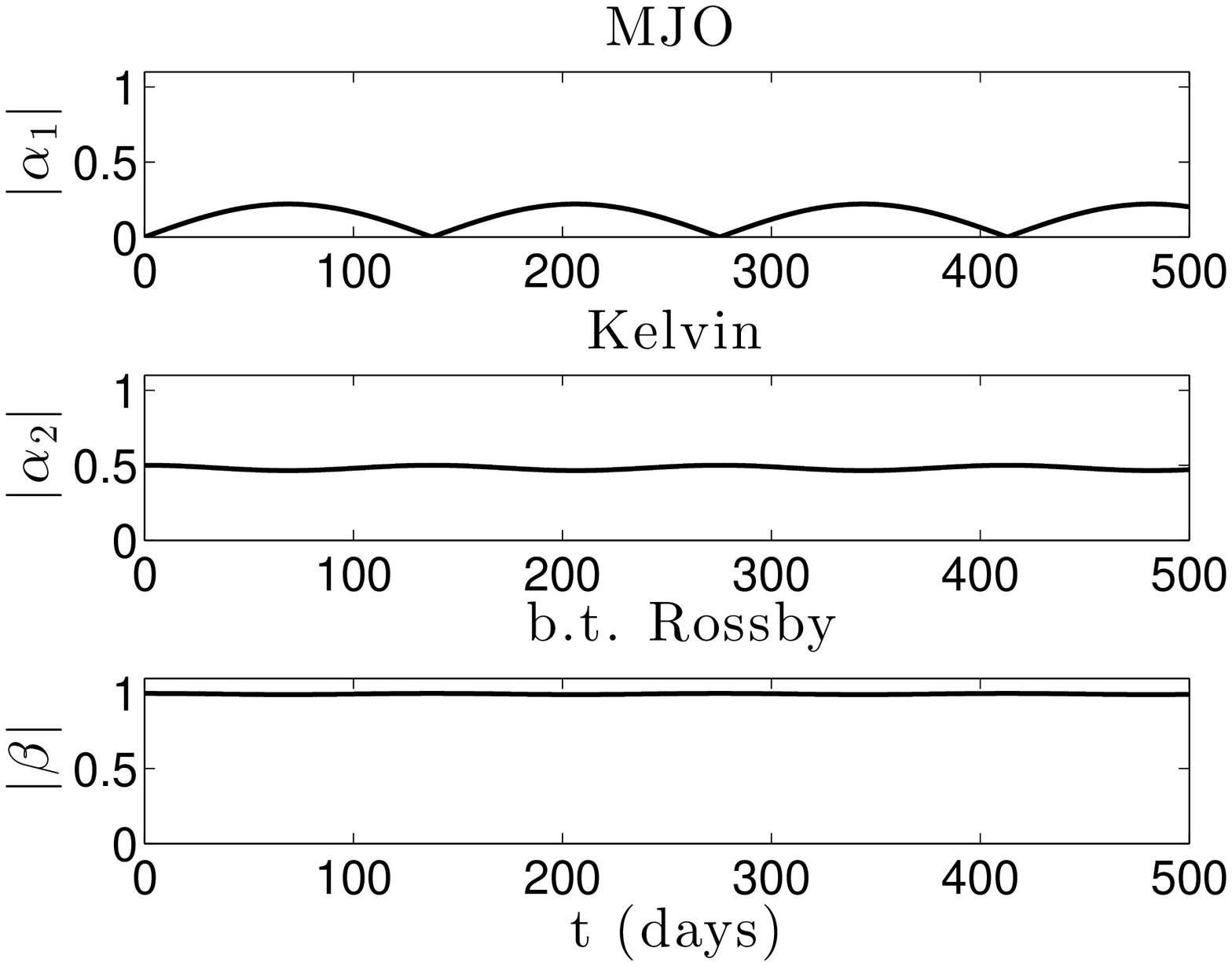}
\includegraphics[width = .45\hsize]{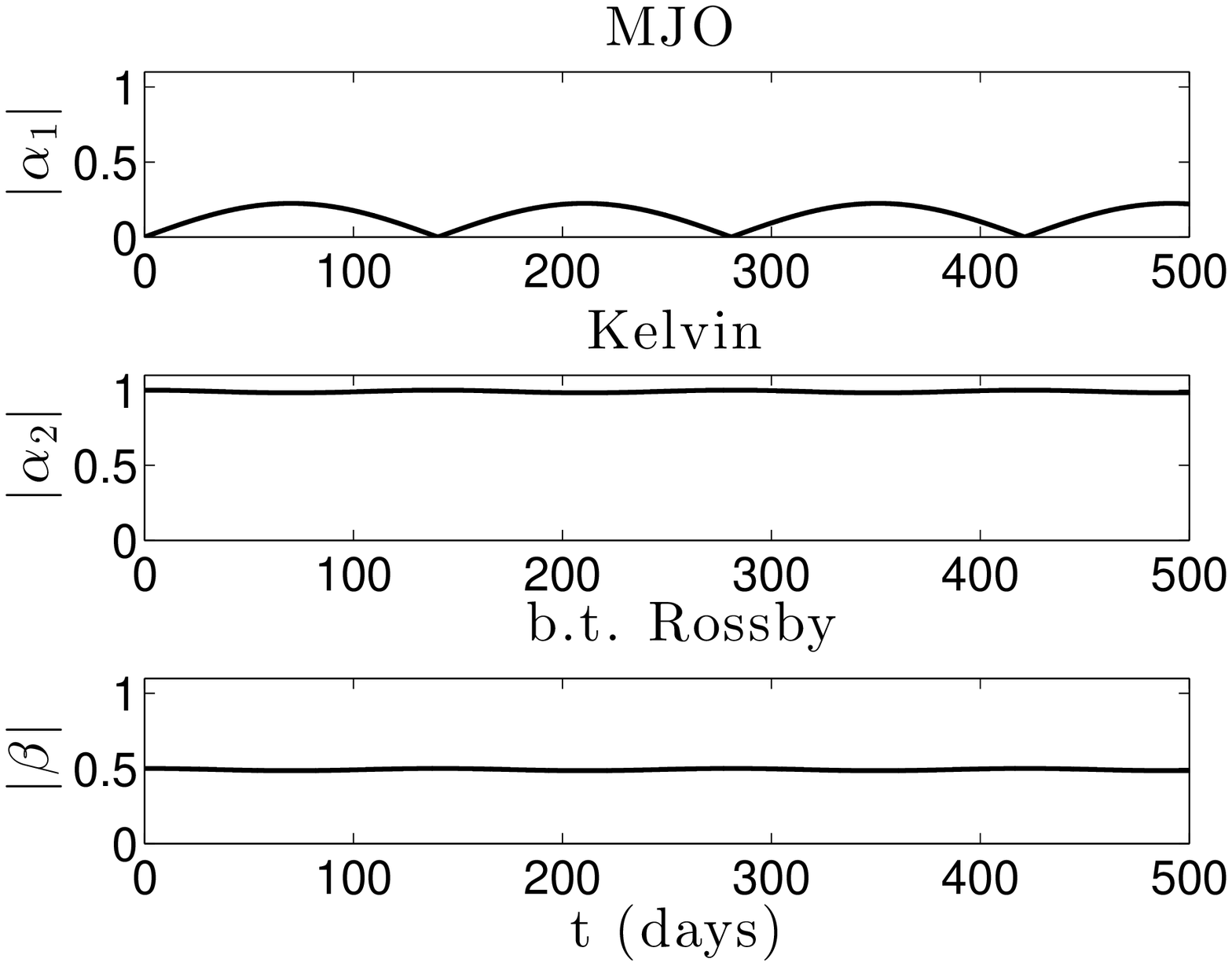}
\hrule
\includegraphics[width = .45\hsize]{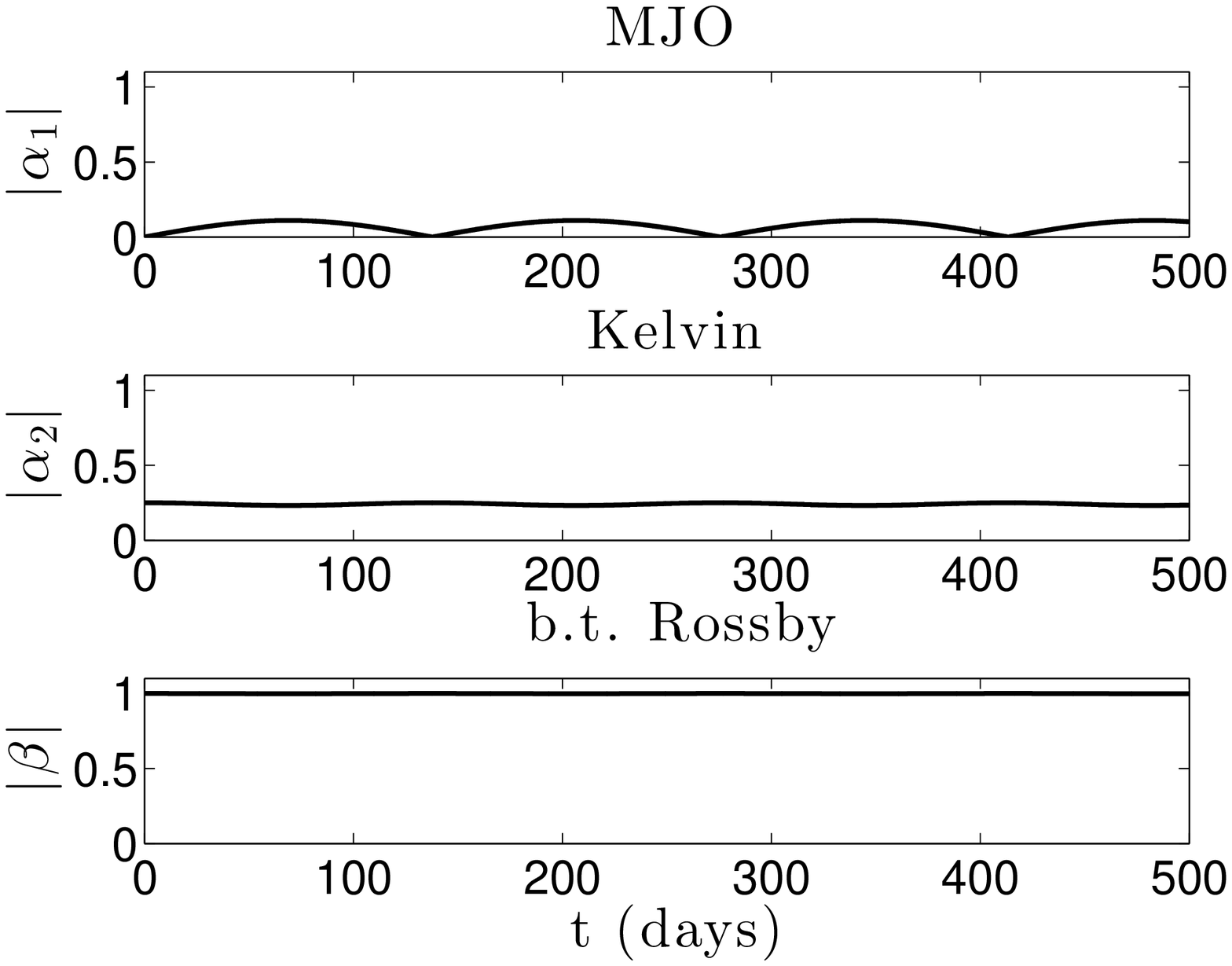}
\includegraphics[width = .45\hsize]{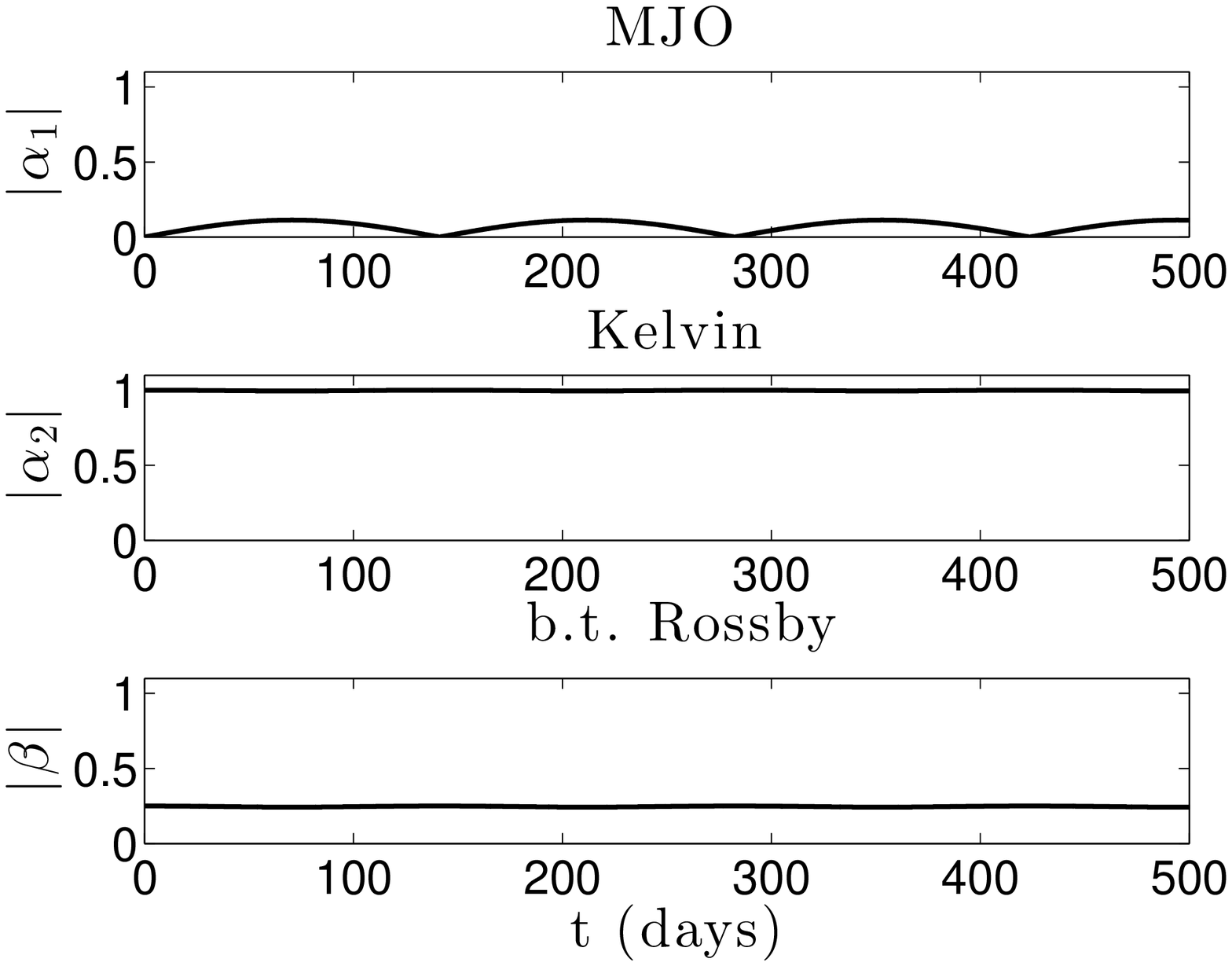}
\caption
{MJO--Kelvin--barotropic Rossby wave interactions.
Initial condition: $\aji$ = 0.
 (a) $|\ajt_{(t=0)}| =0.5$,$ |\beta_{(t=0)}|= 1$;
 (b) $|\ajt_{(t=0)}| =1$, $|\beta_{(t=0)}|= 0.5$;
 (c) $|\ajt_{(t=0)}| =0.25$, $|\beta_{(t=0)}|= 1$;
 (d) $|\ajt_{(t=0)}| =1$, $|\beta_{(t=0)}|= 0.25$.
}
\label{fig_MKB_rdiff}
\end{center}
\end{figure}

Finally,
the sensitivity of the results is illustrated in
Figure~\ref{fig_MKB_rdiff},
which shows numerical results with different initial amplitudes.
As in the MRB case, the amount of energy 
extracted by the MJO varies roughly proportionally to
the amount of energy present initially in the other two modes.

\section{Summary and conclusions}
A new model has been proposed here to describe
MJO initiation and termination and tropical-extratropical interactions.
The model involves the integration of
the barotropic and equatorial baroclinic modes together with
moisture and convective activity envelope,
all interacting together with a conserved energy.
Using the method of multiscale asymptotics with multiple time scales,
simplified asymptotic equations were derived for 
the resonant interaction of tropical and extratropical waves.
The simplified model is an ODE system for wave amplitudes,
including quadratic nonlinearity as in many traditional three-wave-resonance
equations, and also including cubic self-interaction terms
that arise from nonlinear interactions of water vapor and convective activity.

Example simulations of the ODE system are shown to
illustrate some cases of MJO initiation.
In this model, the MJO is shown to initiate mainly by extracting energy
from either the dry Kelvin wave or the dry equatorial baroclinic Rossy wave.
While the MJO extracts a smaller amount of energy from the
barotropic Rossby waves, the barotropic Rossby waves are
essential for MJO initiation, and the strength of the ensuing
MJO depends on the strength of the barotropic Rossby waves present.
In this way, the barotropic Rossby wave acts as a catalyst for the interaction of
the other two waves.
In other scenarios, it is possible that 
the barotropic Rossby wave may lose or gain energy through wave
interactions if more realistic setups are used.
For example,
(i) zonally varying climatological mean state,
and (ii) shear could affect the interaction mechanisms.
In the future, it would also be interesting to include the effects of
additional meridional modes that are asymmetric with respect to the equator
in order to model the Boreal summer MJO and 
monsoon intraseasonal variability
\cite{tms15}.
The authors are currently pursuing these issues and will report on them in the near future.

The simplified models suggest specific classes of waves that could be investigated
in observations among the plethora of possible waves and wave interaction scenarios.
More specifically,
one can imagine wave interactions among many different types of waves.
Here, in this paper, we have identified specific classes of three-wave interactions
involving the MJO.
For example,
one class involved a wavenumber-1 MJO, wavenumber-3 dry baroclinic Rossby wave,
and wavenumber-2 barotropic Rossby wave.
It would be interesting to identify this specific class of waves in observational
data and investigate their interactions.
Such an investigation and comparison with the model could be challenging
for many reasons;
for example, as described above,
the mean state of the model version here does not include a zonally
varying climatological mean state nor wind shear,
which may influence the wave interactions.

%Various observational analysis 
%argues that the MJO can be triggered 
%by extratropical Rossby waves propagating towards equator (\cite{kw92b, mk99}).
%Our results suggest that a possible excitation mechanism 
%in which the intraseasonal variability is associated with 
%energy exchanges involving modes of different resonant triads of equatorial waves.
%Here the barotropic Rossby waves appear as
%the conveyor belt for energy transfer between the MJO and equatorial waves. 
%This property of the barotropic Rossby waves were also suggested in the models by
%in \cite{rd06, rd09, rtm08}.
%We have not found any observational study
% investigating the energy of extratropical Rossby waves. 
%On the other hand, some important physical features may change the outcome
%from the resonance condition. 
%The authors are currently investigating the effect of
%Walker circulation result from the imbalanced heating and moisture sources.
% In this case, the Walker circulation enables the direct energy exchange
% between the MJO and the barotropic Rossby waves. 
%Another example is that, 
%the effects of climatological mean wind shear have been neglected here,
%although it can have a significant impact on 
%tropical--extratropical interactions~\cite{hj91,mb03, w72, w81, w82, wx96}.
%In the future, it would also be interesting to include the effects of
%additional meridional modes that are asymmetric with respect to the equator
%in order to model the Boreal summer MJO and 
%monsoon intraseasonal variability
%\cite{tms15}.
%The authors are currently pursuing these issues and will report on them in the near future.
%

\appendix 
%\numberwithin{equation}{section}

\section{The explicit formulation of the meridional truncated system}\label{sec_trun_sys}
Applying the meridional truncation of (\ref{merid_trunc})
to the long-wave-scaled system of (\ref{eq_long}) 
leads to the \textbf{truncated system}:
\begin{subequations}
\begin{align}
& L^2 YB_{t'} - Y B_{x'}= \CC_{\text{T}}+\delta^2 \left(Y B_{x'x't'}+\DD_{\text{T}}\right) \\
& \lo_{t'} -\lo_{x'} + \vi + \frac{1}{\sqrt{2}}\delta \Hbar \ao = \CC_{\lo}  \\
&\lt_t-\lt_x = \CC_{\lt}\\
&\ro_{t'} +\ro_{x'} + \frac{1}{\sqrt{2}}\delta \Hbar \ao  =\CC_{\ro}\\
 &\rt_{t'}+\rt_{x'}-\sqrt{2} \vi'  = \CC_{\rt}\\
&  \sqrt{2}\rt - \lo =\delta^2(-\vi'_{t'}  +\DD_{\vi})\\
& \qo_{t'} + \frac{\tildeq}{\sqrt{2}}(\ro_{x'}-\lo_{x'}) + \frac{\tildeq}{\sqrt{2}}\vi' +\delta \Hbar \ao = \CC_{\qo}\\
& \qt_{t'} + \frac{\tildeq}{\sqrt{2}}(\rt_{x'}-\lt_{x'})-\tildeq \vi' =\CC_{\qt}\\
& \ao_{t'} -\Gamma \abar \qo = c_1 \Gamma \ao\qo+ c_{10}\Gamma\ao\qt \label{eq_merid_a}
\end{align} 
\label{eq_merid}
\end{subequations}
which is written in vector form in (\ref{eq_trunc}).
Here $\CC$ and $\DD$ are bilinear terms at different orders:
\begin{subequations}
\begin{align}
&\CC_{\text{T}}= - \frac{c_2}{2} (\ro-\lo)^2_{x'} -\frac{c_3}{2}\left[(\ro-\lo)(\rt-\lt)\right]_{x'} \nonumber \\
&\qquad -\frac{c_4}{2} (\rt-\lt)^2_{x'} -\frac{c_5}{2\sqrt{2}} (\ro-\lo)\vi' - \frac{c_6}{2\sqrt{2}}(\rt-\lt)\vi'   \\
&\CC_{\lo} = -c_2 B_{x'}\lo  -c_8 B_{x'}\lt -2c_2B \lo_{x'} -c_3B \lt_{x'} 
\nonumber\\&\qquad 
+ c_2(\ro-\lo)B_{x'} + \frac{c_3}{2}(\rt-\lt) B_{x'}  -\frac{c_5}{\sqrt{2}}\vi' B \\
&\CC_{\lt} = -{c_9 B_{x'}\lo}  -c_4 B_{x'}\lt  -{ c_3 B \lo_{x'}} -2c_4 B \lt_{x'} 
 \nonumber\\&\qquad
 + \frac{c_3}{2}(\ro-\lo)B_{x'} +c_4 (\rt-\lt) B_{x'} - \frac{c_6}{\sqrt{2}}\vi' B\\
&\CC_{\ro}=-c_2 B_{x'}\ro  -{ c_8 B_{x'}\rt } -2c_2B \ro_{x'} -{ c_3B \rt_{x'}  }\nonumber\\
&\qquad - c_2 (\ro-\lo)B_{x'} -\frac{c_3}{2}(\rt-\lt) B_{x'} + \frac{c_5}{\sqrt{2}}\vi' B   \\
&\CC_{\rt} = -{c_9 B_{x'}\ro}  -c_4 B_{x'}\rt  - c_3 B \ro_{x'} -2c_4 B \rt_{x'}  \nonumber\\
&\qquad- \frac{c_3}{2}(\ro-\lo)B_{x'} -c_4 (\rt-\lt) B_{x'} + \frac{c_6}{\sqrt{2}}\vi' B\\
&\CC_{\qo}=-c_2 B_{x'}\qo  - 2c_2B \qo_{x'}  -c_8 B_{x'}\qt -c_3B \qt_{x'} \\
&\CC_{\qt}=-c_4 B_{x'}\qt  - 2c_4B \qt_{x'}- c_9 B_{x'}\qo-c_3 B \qo_{x'} 
\end{align}
\label{bilinear_c}
\end{subequations}
and 
\begin{subequations}
\begin{align}
&\DD_{\text{T}}= -\frac{c_5}{\sqrt{2}L^2}\left[(\ro-\lo)\vi' \right]_{x'x'} - \frac{c_6}{\sqrt{2}L^2}\left[(\rt-\lt)\vi'\right]_{x'x'} + 2c_7 \vi'^2_{x'}\\
&\DD_{\vi} = -c_7B_{x'}\vi'  -2c_7 B \vi'_{x'} +2c_7 \vi' B_{x'}\nonumber\\
&\qquad+\frac{1}{\sqrt{2}}\left[ \frac{c_5}{L^2}(\ro-\lo)B_{x'x'}+\frac{c_6}{L^2} (\rt-\lt) B_{x'x'}\right]
\end{align}
\end{subequations}

where the coefficients $c_j$s are defined by parabolic cylinder functions:
\begin{align}
\begin{array}{ll}
  c_1 = \int_{-\infty}^{\infty} \Phi_0^3 \mathrm{d}y, 
& c_2 =- \frac{L}{2}\int_{-\infty}^{\infty} \Phi_0^2\cos(Ly) \mathrm{d}y,\\ 
c_3 =- L\int_{-\infty}^{\infty} \Phi_0\Phi_2\cos(Ly) \mathrm{d}y, 
&  c_4 =- \frac{L}{2}\int_{-\infty}^{\infty} \Phi_2^2\cos(Ly) \mathrm{d}y, \\
c_5 =- L^2\int_{-\infty}^{\infty} \Phi_0\Phi_1\sin(Ly) \mathrm{d}y,
&c_6 =- L^2\int_{-\infty}^{\infty} \Phi_1\Phi_2\sin(Ly) \mathrm{d}y, \\
  c_7 =- \frac{L}{2}\int_{-\infty}^{\infty} \Phi_1^2\cos(Ly) \mathrm{d}y,
&c_8 = \int_{-\infty}^{\infty} \Phi_0\Phi_2'\sin(Ly) \mathrm{d}y,\\
c_9 = \int_{-\infty}^{\infty} \Phi_0'\Phi_2\sin(Ly) \mathrm{d}y,
  &c_{10} = \int_{-\infty}^{\infty} \Phi_0^2\Phi_2 \mathrm{d}y, 
\end{array}
\end{align}
Notice that the equality $c_8+c_9=c_3$ helps to conduct the energy conservation.

The truncated system in (\ref{eq_merid}) conserves a total energy 
if the term $\ao\qt$ is neglected in (\ref{eq_merid_a}):
\begin{align}
&\tilde{\mathcal{E}} = \frac{1}{4}\int_0^X  2YL^2(B_x^2+B^2) + \lo^2+\lt^2+\ro^2+\rt^2 + \frac{2\delta^2\Hbar}{\tildeq\Gamma} (\ao-\abar \log(\ao+\abar))\nonumber\\
&\quad +\frac{1}{\tildeq(1-\tildeq)}\left\{\left[\qo-\frac{\tildeq}{\sqrt{2}}(\ro+\lo)\right]^2+\left[\qt-\frac{\tildeq}{\sqrt{2}}(\rt+\lt)\right]^2\right\} \mathrm{d}x 
\end{align}
For this reason,
the term $\ao\qt$ in (\ref{eq_merid_a}) will be neglected
throughout the present paper.

\section{Asymptotic expansion of the truncated system}\label{sec_trun_exp}

Here an explicit expansion is presented of the truncated system from (\ref{eq_merid}), which was written in vector form in (\ref{eq_trunc}).
Using an asymptotic expansion to the third order, 
the \textbf{ansatz} is:
\begin{subequations}
\begin{align}
& B = \delta^2 B_1 + \delta^3 B_2 + \delta^4 B_3 + O(\delta^5)\nonumber\\
& (\lo,\lt,\ro,\rt,\vi', \qo,\qt) \nonumber\\
&\qquad\qquad= \delta^2( \lo_1,\lt_1,\ro_1,\rt_1,\vi_1, \qo_1,\qt_1) \nonumber\\
&\qquad\qquad +\delta^3( \lo_2,\lt_2,\ro_2,\rt_2,\vi_2, \qo_2,\qt_2) \nonumber\\
&\qquad\qquad +\delta^4(\lo_3,\lt_3,\ro_3,\rt_3,\vi_3, \qo_3,\qt_3)+O(\delta^5) \\
& \ao = \delta \ao_1 + \delta^2 \ao_2 + \delta_3\ao_3 + O(\delta^4)
\end{align}
\end{subequations}
All variables are assumed to depend on three time scales:
$t'$ and $T_1=\delta t'$ and $T_2=\delta^2 t'$.

To simplify notation, define 
$\vec{U}_n = ( \lo_n,\lt_n,\ro_n,\rt_n,\vi_n, \qo_n,\qt_n, \ao_n)$, 
and let $\LLU$ denote the spatial linear operator.

Using this ansatz, the truncated system (\ref{eq_merid}), or its vector formulation (\ref{eq_trunc}), can be asymptotically expanded
over three orders of magnitude.
The \textbf{first order system} is
\begin{subequations}
\begin{align}
& L^2 YB_{1t'} - Y B_{1x'}  = 0 ,\\
& \NN \vec{U}_{1t'} +\LLU \vec{U}_1 = 0, 
\end{align}
\label{eq_merid_exp_1}
\end{subequations}
The \textbf{second order system} is
\begin{subequations}
\begin{align}
&  L^2 YB_{2t'} - Y B_{2x'}  = -L^2 YB_{1T_1} ,\\
&  \NN \vec{U}_{2t'} +\LLU \vec{U}_2 = -\vec{U}_{1T_1}+\vec{F}_{\vec{U}2},
\end{align}
\label{eq_merid_exp_2}
\end{subequations}
The \textbf{third order system} is
\begin{subequations}
\begin{align}
&  L^2 YB_{3t'} - Y B_{3x'}  = -L^2 YB_{1T_2} -L^2 YB_{2T_1} +YB_{1x'x't'}+\CC_{1\text{T}},\\
&  \NN \vec{U}_{3t'} +\LLU \vec{U}_3 = -\vec{U}_{1T_2}-\vec{U}_{2T_1}+\vec{F}_{\vec{U}3}.
\end{align} 
\label{eq_merid_exp_3}
\end{subequations}
Here $\NN = \text{diag}(1,1,1,1,0,1,1,1)$ is the $8\times8$ matrix just to eliminate $\partial_{t'}$ for the $\vi$ variable.
Forcing terms $\vec{F}_{\vec{U}2}$ and $\vec{F}_{\vec{U}3}$ are quantities from the lower order:
\begin{subequations}
\begin{align}
&\vec{F}_{\vec{U}2} = (0,0,0,0,0,0,0, c_1 \Gamma \ao_1\qo_1), \\
&\vec{F}_{\vec{U}3}= ( \CC_{1\lo}, \CC_{1\lt}, \nonumber\\
&\qquad\qquad   \CC_{1\ro}, \CC_{1\rt}, -\vi_{1t'},  \CC_{1\qo}, \CC_{1\qt}, c_1\Gamma(\ao_1\qo_2+\ao_2\qo_1 ) ).
\end{align}
\end{subequations}
Here $\CC_1$ denotes the bilinear operations are performed on $\vec{U}_1$.
Like the 2D asymptotic expansion, 
system (\ref{eq_merid_exp_1})--(\ref{eq_merid_exp_3}) carries
similar approximated energy conservation that 
\begin{equation}
\frac{\mathrm{d}}{\mathrm{d}t'}\tilde{\mathcal{E}} = o(\delta^4).
\end{equation}

\section{Details of the auxiliary problem}\label{sec_aux}
In this appendix, the auxiliary problem is written explicitly in
terms of the new variables
$\vec{W} = (K, R, Q, A, \vi, \chi),$
where the change of variables was described abstractly in
section~\ref{sec_aux_new}.

\subsection{Reformulation of the auxiliary problem}
Equation~(\ref{aux_abs_tilde}) can be concretely written as
\begin{subequations}
 \begin{numcases}{}
 K_t + K_x + \frac{1}{\sqrt{2}}\Hbar A = F_{\ro} \\
  R_t + \sqrt{2}\chi_x -\sqrt{2}\vi +  \Hbar A = \sqrt{2}F_{\lo}+2 F_{\rt} \\
  Q_t + \frac{\tildeq}{\sqrt{2}}\left(K_x-\frac{1}{2\sqrt{2}}R_x + \frac{1}{2}\chi_x\right)+\frac{\tildeq}{\sqrt{2}}\vi +  \Hbar A =  F_{\qo} \\
  A_t -  \Gamma \abar Q = F_{\ao} 
\end{numcases} 
\begin{numcases}{}
 0\cdot \vi _t + \chi = F_{\vi} \label{krqa_v} \\
 \chi_t + \frac{1}{\sqrt{2}} R_x - 3\vi - \frac{1}{\sqrt{2}}\Hbar A = -F_{\lo}+\sqrt{2}F_{\rt} \label{krqa_chi}
\end{numcases}
\begin{numcases}{}
\lt_t - \lt_x = F_{\lt}\\
\qt_t + \frac{\tildeq}{\sqrt{2}}(\frac{1}{4}R_{x'} + \frac{\sqrt{2}}{4}\chi_{x'}-\lt_{x'})-\tildeq \vi'  = F_{\qt}
\end{numcases}
\end{subequations}
The equations (\ref{krqa_v})-(\ref{krqa_chi}) are used to eliminate $\vi$ and $\chi$ in the other equations:
\begin{subequations}
\begin{align}
&\chi = F_{\vi}\\
&\vi = \frac{1}{3\sqrt{2}}\left(R_x-\Hbar A\right)+ \frac{1}{3}\left(F_{\vi ,t} +F_{\lo}-\sqrt{2}F_{\rt}\right)
\end{align}
\end{subequations}
so that the system becomes
\begin{subequations}
\begin{numcases}{}
 K_t + K_x + \frac{1}{\sqrt{2}}\Hbar A = F_{\ro} \label{krqa_k} \\
 R_t + \frac{1}{3}R_x + \frac{4}{3} \Hbar A =\nonumber\\
 \quad\qquad \frac{4\sqrt{2}}{3}F_{\lo}+\frac{4}{3} F_{\rt} -\sqrt{2} F_{\vi, x}+\frac{\sqrt{2}}{3}F_{\vi,t} \label{krqa_r} \\
 Q_t + \frac{\tildeq}{\sqrt{2}}K_x-\frac{\tildeq}{12}R_x+ (\frac{\tildeq}{6}-1) \Hbar A =\nonumber\\
 \qquad \quad -\frac{\tildeq}{3\sqrt{2}}F_{\lo} +\frac{\tildeq}{3} F_{\rt} + F_{\qo}-\frac{\tildeq}{2\sqrt{2}}F_{\vi,x}-\frac{\tildeq}{3\sqrt{2}}F_{\vi,t} \label{krqa_q}\\
 A_t -\Gamma \abar Q = F_{\ao}  \label{krqa_a}
\end{numcases}
\begin{numcases}{}
 \lt_t - \lt_x = F_{\lt} \label{krqa_lt}\\
 \qt_t-\frac{\tildeq}{12\sqrt{2}}R_x - \frac{\tildeq}{\sqrt{2}}\lt_x + \frac{\tildeq\Hbar}{3\sqrt{2}}A = \nonumber\\
\qquad\qquad F_{\qt}+\frac{\tildeq}{3}F_{\lo}-\frac{\sqrt{2}\tildeq}{3}F_{\rt}-\frac{\tildeq}{4}F_{\vi,x}- \frac{\tildeq}{3\sqrt{2}}F_{\vi,t} \label{krqa_qt}
\end{numcases}
\label{krqa}
\end{subequations}
When $\FFW=0$, the system (\ref{krqa}), (\ref{krqa_k})-(\ref{krqa_a}) forms the KRQA system
as in~\cite{ms09pnas}, while the equation (\ref{krqa_lt}) for $\lt$ is independent, 
and equation (\ref{krqa_qt}) for $\qt$  is slaved
by KRQA system and $\lt$.

\subsection{Forcing vector for the auxiliary problem}
Recall the earlier presentation of auxiliary problems in
(\ref{eq_merid_exp_1})--(\ref{eq_merid_exp_3})
in terms of the variables $\vec{U}$.
In the transformation from $\vec{U}$ to $\vec{W}$
that was described abstractly in section~\ref{sec_aux_new},
the forcing vector $\FFU$ is transformed into forcing vector $\FFW$.
The transformation of the forcing vector is now presented explicitly
for each of the cases in
(\ref{eq_merid_exp_1})--(\ref{eq_merid_exp_3}).
Using the notation
$$B= \delta^2B_1+\delta^3B_2 +\delta^4B_3 + O(\delta^5),$$
 and 
 $$\vec{W} =\delta^2\vec{W}_1+\delta^3\vec{W}_2 +\delta^4\vec{W}_3 + O(\delta^5) ,$$
the \textbf{leading order system} is
\begin{subequations}
\begin{align}
& L^2 YB_{1yyt'} - Y B_{1x'}= 0\\
& \vec{W}_{1t'} + \LLW \vec{W}_1 = 0
\end{align}
\label{krqa_1}
\end{subequations}
The \textbf{second order system} is
\begin{subequations}
\begin{align}
& L^2 YB_{2yyt'} - Y B_{2x'}=  -B_{1yyT_1},\\
& \vec{W}_{2t'} + \LLW \vec{W}_2 =  -\vec{W}_{1T_1}+ \vec{F}_{\vec{W}2},
\end{align}
\label{krqa_2}
\end{subequations}
and the \textbf{third order system} is
\begin{subequations}
\begin{align}
& L^2 YB_{3yyt'} - Y B_{3x'}=  -B_{1yyT_2}-B_{2yyT_1}+B_{1x'x't'}+\CC_{1\text{T}}\\
& \vec{W}_{3t'} + \LLW \vec{W}_3 =  -\vec{W}_{1T_2}-\vec{W}_{2T_1}+ \vec{F}_{\vec{W}3}.
\end{align}
\label{krqa_3}
\end{subequations}
Here 
\begin{subequations}
\begin{align}
& \vec{F}_{\vec{W}2} = (0,0,0,c_1\Gamma A_1Q_1, 0,0)\\
 & \vec{F}_{\vec{W}3} = (\CC_{1\ro}, 
 \frac{4\sqrt{2}}{3}\CC_{1\lo}+\frac{4}{3}\CC_{1\rt}+\sqrt{2}\vi_{1x't'}-\frac{\sqrt{2}}{3}\vi_{1t't'}, \nonumber\\
 &\qquad
 -\frac{\tildeq}{3\sqrt{2}}\CC_{1\lo}+\frac{\tildeq}{3}\CC_{1\rt}+\CC_{1\qo}+\frac{\tildeq}{2\sqrt{2}}\vi_{1x't'}+\frac{\tildeq}{3\sqrt{2}}\vi_{1t't'}, \nonumber\\
 &\qquad
 c_1\Gamma(A_1Q_2+A_2Q_1 )+\CC_{1\ao}, \CC_{1\lt},\nonumber\\
 &\qquad 
  \CC_{1\qt}+\frac{\tildeq}{3}\CC_{1\lo}-\frac{\sqrt{2}\tildeq}{3}\CC_{1\rt}+\frac{\tildeq}{4}\vi_{1x't'}+\frac{\tildeq}{3\sqrt{2}}\vi_{1t't'} )
\end{align}
\end{subequations}
Note that for the first and second order, $\chi_1=\chi_2 =0$, and 
\begin{equation}
\vi_{n} = \frac{1}{3\sqrt{2}}\left(R_{nx'} - \Hbar A_n \right), \quad n = 1, 2.
\label{eq_v12}
\end{equation}

\section{Details for multiscale analysis }\label{sec_multi}
\subsection{Solving linear system with forcing} \label{sec_forcing}
Here a brief summary for solving linear system with forcing is provided,
which is a guideline of the multiscale analysis. 
Readers may find information from other references, e.g., \cite{m03}.
Let $\vec{W}$ solve the linear system with forcing:
\begin{equation}
\partial_{t'} \vec{W}+ \LLW \vec{W} =\hat{ \vec{F}}e^{i(kx - ct')}, \vec{W}|_{t'=0}=0.
\label{eq_forcing}
\end{equation}
Next, assume that $\vec{W}$ is the superposition of eigenmodes:
\begin{equation}
\vec{W}(x',t') = \sum_{s} a_s(t')\vec{r}_s(k) e^{i(kx'-\omega_s(k)t')}.
\label{eq_Wset}
\end{equation}
Two cases might happen:
\begin{enumerate}
\item If $\omega_s(k)\neq c$, then $a_s = \frac{1}{\omega_s(k)-c} \left(e^{i(\omega_s(k)-c)t'}-1\right)\left(\vec{l}_s \cdot \hat{\vec{F}}\right) $,
\item If $\omega_s(k)=c$, then $a_s = t \left(\vec{l}_s \cdot \hat{\vec{F}} \right)$.
\end{enumerate}
At the second case, when $\vec{l}_s \cdot \hat{\vec{F}} \neq 0$, linear growth in time will happen.
Therefore the solution $\vec{W}$ is bounded if either
1) $\omega_s(k)\neq s$ for all s;
or 2) $\omega_s(k)=c$ and $\vec{l}_s\cdot \hat{\vec{F}}=0$.
The above is the principal of the multi-scale analysis.

 \subsection{Solutions for the second order equations} \label{sec_2nd}
At the second order, 
the only nonlinear term is in the $A$ equation in (\ref{krqa_2}).
The barotropic and baroclinic modes are still decoupled.
They are studied separately.
For the baroclinic wave, the only nonlinear term does not 
generate resonance condition between different eigenmodes.
Therefore, considering one mode is sufficient to study solutions for
the second order equations.

Assume that the leading order solution has a form of
$$\vec{W}_1 =
 \aji(T_1,T_2) e^{i\thji} \vec{r}_{j_1}+\text{C.C.}$$

Here $\thji = \kji x' - \oji_{(\kji)}t'$ is the phase, and 
$$\vec{r}_{j_1} = (\hat{K}_{1,1},\hat{R}_{1,1},\hat{Q}_{1,1},\hat{A}_{1, 1},\hat{\lt}_{1,1},\hat{\qt}_{1,1})$$ is the eigenvector for the $k=\kji$ eigenmode of certain wave type (e.g., Kelvin, MJO, moist Rossby, dry Rossby).

Also assume that the leading order baroclinic wind is
\begin{align}
W_1 &= \aji e^{i\thji} \vec{r}_{1} + \ajt e^{i\thjt} \vec{r}_{2}+ \text{C.C.},\nonumber
\end{align}
where $\theta_{j} = k_j x' + \omega_j t' $, and 
$$\vec{r}_{j} = (\hat{K}_{1,j},\hat{R}_{1,j},\hat{Q}_{1,j},\hat{A}_{1, j},\hat{\lt}_{1,j},\hat{\qt}_{1,j}), \quad j = 1, 2. $$
Here the eigenvectors are chosen so that $\hat{K}_{1,j}$s are real numbers.
The eigenvector $\vec{r}_j$ are normalized by their energy unit, so that 
\begin{align}
&E_j = \vec{r}_j^{\dag} \mathcal{H} \vec{r}_j=1, \quad j = 1, 2,
\end{align}
where $\mathcal{H}$ is the Hessian matrix of the conserved energy for the linear system.

At the second order, the forcing terms come from the leading order (\ref{krqa_2}).
The solution consists of two parts: 
the homogeneous solution and the nonhomogeneous solution from the forcing terms.
The homogeneous solution are just linear eigenmodes which is considered to be absorbed into the leading order.
Therefore the second order solution is completely determined by forcing terms:
\begin{equation}
\vec{F} = -\aji \partial_{T_1}\vec{W}_{1}-\alpha_1^*\partial_{T_1} \vec{W}^*+
		\left[
		 \begin{array}{c}
		0\\
		0\\
		0\\
		F_3\\
		0\\
		0
		 \end{array}
		\right],
\end{equation}
where 
$$F_3 =c_1 \Gamma  \left( \aji^2 e^{i2\thji} \hat{A}_{1, j_1} \hat{Q}_{1, j_1} +\aji \alpha^*_1 \hat{A}_{1, 1}^* \hat{Q}_{1, j_1} \right)+\text{C.C.}$$
 The right-hand-side of the second order has three phases: 
 $$e^{\pm i\thji}, e^{\pm i2\thji}, \text{ and } e^{0}.$$
  \begin{enumerate}
 \item Phase $e^{\pm i\thji}$ \\
These two phases generate secular growth. 
The phase $e^{i\thji}$requires that 
 $$\vec{l}_{1 (\kji)} \cdot \hat{\vec{F}}_{(\kji)}= \partial_{T_1}\aji = 0.$$
The same thing applies for phase $e^{-i\thji}$, so that $  \partial_{T_1}\aji=0$.

 \item Phase $e^{\pm i2\tho}$\\
 This part comes from the Q-A nonlinear term. 
 Because it does not generate secular growth,
 the system can be solved and the solution is
 \begin{align}
 & \hat{\vec{W}}_{2(2\kji, t',T_2)} = \nonumber\\
  &\quad \mathcal{R}_{(2\kji)} (e^{-2i\oji_{(\kji)}t} \mathcal{I}-e^{-\mathcal{D}(2\kji) t}) (\mathcal{D}_{(2\kji)} -2i\oji_{(\kji)}\mathcal{I})^{-1} \mathcal{L}_{(2\kji)} \hat{\vec{F}}_{(2\kji)}
  \end{align}
 
 \item Phase $e^0$\\
 This part also stems from the Q-A nonlinear term.
 Although it has $0$-frequencies, which may resonant with the MJO, moist Rossby mode, 0-frequency mode, fast west-propagating $\lt$ mode,
 but it does not happen because
 $$\vec{l}_{\text{MJO}(0)}  \cdot \hat{\vec{F}}_{(0)} = \vec{l}_{\text{mR}(0)} \cdot \hat{\vec{F}}_{(0)} = \vec{l}_{\text{0fr}(0)}  \cdot \hat{\vec{F}}_{(0)} = \vec{l}_{\lt (0)} \cdot \hat{\vec{F}}_{(0)} = 0,$$
which is also stated in ~\cite{m03}.
 The phase $e^0$ part have the contribution to the solution as
 $$ \hat{\vec{W}}_{2(k=0,t',T_2) }= \mathcal{R}_{(0)} 
\left( \text{diag} \left[
 \begin{array}{c}
 \frac{1}{i\okk_{(0)}}\\
 0 \\
 0\\
 0\\
 \frac{1}{i\odr_{(0)}}\\
  0
 \end{array}
 \right]\right)
 (\mathcal{I}-e^{-\mathcal{D}_{(0)}t'} )\mathcal{L}_{(0)}\hat{\vec {F}}_{(0)}
 $$
 \end{enumerate}
 
 One remark is that because the leading order as forcing terms does not have $T_1$ dependence,
the second order solution, 
 is also independent of $T_1$,
 i.e.,
 $$\frac{\partial}{\partial T_1}\vec{W}_2 = \vec{0}.$$

\subsection{Resonance condition for the third order equation}
At the third order, there are nonlinear terms mixing baroclinic and barotropic terms,
together with the nonlinear $q$-$a$ interaction.

To simplify mathematical calculations,
 the $\vec{W}_1$ variables are transformed back to the $\vec{U}_1$ variables,
 where the Fourier coefficients can be written as
\begin{subequations}
\begin{align}
&\hat{\lo}_{1,j} =  \frac{1}{2\sqrt{2}}\hat{R}_{1,j} , \quad \hat{\rt}_{1,j} = \frac{1}{4}\hat{R}_{1,j}, \nonumber\\
&\hat{\qo}_{1,j} = \hat{Q}_{1,j}, \quad \hat{\ao}_{1,j} = \hat{A}_{1,j}, \nonumber\\
&\hat{\vi}_{1,j} = \frac{1}{3\sqrt{2}}\left(i\kji \hat{R}_{1,j} - \Hbar \hat{A}_{1,j} \right), \quad j = 1,2. \nonumber
\end{align}
\end{subequations}

Here the three groups of terms in (\ref{eq_3wv_ode}) are explained separately:

\begin{enumerate}
\item Cubic terms.

 The cubic terms in (\ref{eq_3wv_ode}) are associated with 
 the product of seconder order and the first order from the nonlinear $q$-$a$ interaction. 
These terms are self-interaction -- they do not interact with other eigenmodes.

Let $(K_2, R_2,Q_2,A_2,\lt_2,\qt_2)$ be the solution obtained in Section~\ref{sec_2nd}.
The term $c_1\Gamma(A_1Q_2+A_2Q_1)$ contributes to the resonant condition
by the nonlinear interactions between the first and second order.
 The leading order has phases $e^{\pm i(\kji x' -\oji(\kji)t')}$,
 whereas the second order has phases $e^{\pm i(2 \kji x' -2\oji(\kji)t')}$, $e^{\pm i (i 2\kji x' -\omega(2\kji)t' )}$, $e^{\pm i \okk(0)t'}$, $e^{\pm i \odr(0)t'}$, and $e^0$.
 The product of the phases $e^{\pm i (2\kji x' - 2\oji(\kji)t')}$ and $e^0$ in the second order together with 
 the phases $e^{\pm i(\kji x' -\oji(\kji)t')}$ at the leading order will contribute to secular growth. 
 
 The coefficient $id_4$ can be written as
%\begin{subequations}
\begin{align}
 d_4 &= \vec{l}_{1} \cdot (0,0,0,d_{44},0,0  ) \label{eq_d4}
 \end{align} 

where 

\begin{align}
d_{44} &= \hat{A}_{1,1(-\kji)}\hat{Q}_{2,1(2\kji)} +\hat{Q}_{1,1(-\kji)}\hat{A}_{2,1(2\kji)} \nonumber\\
& + \hat{A}_{1,1(\kji)}\hat{Q}_{2,1(0)}+\hat{Q}_{1,1(\kji)}\hat{A}_{2,1(0)}
\end{align}

where $\vec{l}_{1}$ is the left eigenvector of the choice of eigenmode at the leading order, 
and $\hat{\vi}_{1,1}$ is in terms of $\hat{R}_{1,1}$ and $\hat{A}_{1,1}$ from (\ref{eq_v12}),
that
$$\hat{\vi}_{1,1} = \frac{1}{3\sqrt{2}}\left(i\kji \hat{R}_{1,1} - \Hbar \hat{A}_{1,1} \right).$$
And $id_7$ in the equation for $\ajt$, same calculations are applied, but starting from 
another eigenmode.

\item
Linear terms 

Linear terms come from dispersive effect in long-wave scaled system.
These are also self-interaction terms.

For the barotropic part, the coefficient for the linear term $id_2$ writes
$$i d_2 = \frac{1}{L^2}i \kbt^2 \obt.$$
For the baroclinic part, the coefficient $id_5$ writes
\begin{align}
 i d_5&=\vec{l}_{1} \cdot (0,d_{52}, d_{53}, 0,0, d_{56} ),
       \label{eq_d5}
 \end{align} 
where
\begin{align}
d_{52} &=\sqrt{2}\kji\oji\hat{\vi}_{1,1}+\frac{\sqrt{2}}{3}\oji^2\hat{\vi}_{1,1},  \\
 d_{53}&=  \frac{\tildeq}{2\sqrt{2}}\kji\oji\hat{\vi}_{1,1}-\frac{\tildeq}{3\sqrt{2}}\oji^2\hat{\vi}_{1,1}, \\
 d_{56}&=\frac{\tildeq}{4}\kji\oji\hat{\vi}_{1,1}-\frac{\tildeq}{3\sqrt{2}}\oji^2\hat{\vi}_{1,1}.
\end{align}

\item Quadratic terms

The quadratic terms are associated with nonlinear baroclinic--barotropic interactions.

In (\ref{eq_3wv_ode}), coefficient $i d_3$ can be written as 
\begin{equation}
i d_3 = -\frac{1}{\sqrt{E_{\text{T}}}} d_{31}
\end{equation}
where
\begin{align}
 d_{31} =
& - i(\kji+\kjt) 2c_2(\hat{\ro}_{1,1}-\hat{\lo}_{1,1})(\hat{\ro}_{1,2}-\hat{\lo}_{1,2}) \nonumber\\
		&- i(\kji+\kjt) 2c_4(\hat{\rt}_{1,1}-\hat{\lt}_{1,1})(\hat{\rt}_{1,2}-\hat{\lt}_{1,2}) 
\nonumber\\
&- {i c_3(\kji+\kjt)} (\hat{\ro}_{1,1}-\hat{\lo}_{1,1})(\hat{\rt}_{1,2}-\hat{\lt}_{1,2})\nonumber\\
& - {i c_3(\kji+\kjt)} (\hat{\ro}_{1,2}-\hat{\lo}_{1,2})(\hat{\rt}_{1,1}-\hat{\lt}_{1,1})   \nonumber\\
& -\frac{c_5}{\sqrt{2}}\left[(\hat{\ro}_{1,1}-\hat{\lo}_{1,1})\hat{\vi}_{1,2}+(\hat{\ro}_{1,2}-\hat{\lo}_{1,2})\hat{\vi}_{1,1}\right]  \nonumber\\
& -\frac{c_6}{\sqrt{2}}\left[(\hat{\rt}_{1,1}-\hat{\lt}_{1,1})\hat{\vi}_{1,2}+(\hat{\rt}_{1,2}-\hat{\lt}_{1,2})\hat{\vi}_{1,1}\right]  \nonumber\\
\end{align}
and for baroclinic mode, coefficient $i d_6$ can be written as 
\begin{align}
- i d_6 =& \frac{1}{\sqrt{E_{\text{T}}}}\vec{l}_{1} \cdot
 (\hat{F}_{\ro_{2}}, 
 \frac{4\sqrt{2}}{3}\hat{F}_{\lo_{2}}+\frac{4}{3}\hat{F}_{\rt_{2}},
 -\frac{\tildeq}{3\sqrt{2}}\hat{F}_{\lo_{2}}+\frac{\tildeq}{3}\hat{F}_{\rt_{2}}+\hat{F}_{\qo_{2}},
\nonumber\\
 &\qquad
  \hat{F}_{\ao_{2}}, 
 \hat{F}_{\lt_{2}}, 
 \frac{\tildeq}{3}\hat{F}_{\lo_{2}}-\frac{\sqrt{2}\tildeq}{3}\hat{F}_{\rt_{2}}+\hat{F}_{\qt_{2}}
 ),
\end{align}
where $\vec{l}_{1}$ is the left eigenvector for the $j=1$ baroclinic mode,
and $\hat{F}$ are Fourier coefficients of interactions between $j=2$ baroclinic mode and barotropic wind,
They corresponds to bilinear terms $\CC$s in (\ref{bilinear_c}).
One example is given for $\hat{F}_{\lo_{2}}$:
\begin{align}
\hat{F}_{\lo_{2}} =& -i\kbt \left(c_2 \hat{\lo}_{1,2} +c_8 \hat{\lt}_{1,2} \right)
 - i \kjt \left(2c_2\hat{\lo}_{1,2} +c_3\hat{\lt}_{1,2}\right)\nonumber\\
& +i\kbt \left[c_2(\hat{\ro}_{1,2}-\hat{\lo}_{1,2}) +\frac{c_3}{2}(\hat{\rt}_{1,2}-\hat{\lt}_{1,2})  \right] 
%\nonumber\\ & 
-\frac{c_5}{\sqrt{2}}\hat{\vi}_{1,2}
\end{align}
The coefficient $d_9$ is analogous to $d_6$ by swapping $j=1$ and $j=2$.
\end{enumerate}
\section*{acknowledgement}
The research of A.J.M. is partially supported by
Office of Naval Research grant ONR MURI N00014-12-1-0912.
The research of S.N.S. is partially supported by
Office of Naval Research grant ONR MURI N00014-12-1-0912,
ONR Young Investigator Award ONR N00014-12-1-0744,
and National Science Foundation grant NSF DMS-1209409.
S.C. is supported as a postdoctoral research associate by the ONR grants.

\bibliographystyle{plain}
\bibliography{mjobib.bib}{}

\end{document}